\documentclass[]{aa}
\usepackage{txfonts}
\usepackage{graphicx}

\newcommand{\fan}[1]{\phantom{#1}}
\newcommand{\hii}{\ion{H}{ii}}
\newcommand{\mhe}{\mathrm{He}}
\newcommand{\md}{\mathrm{d}}

\begin{document}

\title{Dust and Nebular Emission. I. Models for Normal Galaxies.}

\author{P. Panuzzo\inst{1}\fnmsep\thanks{\email{panuzzo@sissa.it}}
\and A. Bressan\inst{2,1} \and G. L. Granato\inst{2} \and
L. Silva\inst{3} \and L. Danese\inst{1}}

\institute{Scuola Internazionale Superiore di Studi Avanzati,
Via Beirut 4, I-34014 Trieste, Italy
\and
INAF, Osservatorio Astronomico di Padova, Vicolo
dell'Osservatorio 5, I-35122 Padova, Italy
\and
INAF, Osservatorio Astronomico di Trieste, Via G. B. Tiepolo
11, I-34131 Trieste, Italy}

\date{Received ; accepted }

\abstract{We present a model for nebular emission in
star forming galaxies, which takes into account the effects of dust
reprocessing. The nebular emissions (continuum emission, 54 H and 
He recombination lines and 60 nebular lines from UV to IR) have been 
computed
with CLOUDY and then included into GRASIL, our spectrophotometric code
specifically developed for dusty galaxies. The interface between
nebular emission and population synthesis is based on a set of
pre-computed \hii\ region emission models covering a wide range of
physical quantities (metallicity, density, geometry and number of
\ion{H}{i}, \ion{He}{i} and \ion{O}{ii} ionizing photons). These
quantities are fully adequate to describe the emission properties of
the majority of star-forming and starburst galaxies. 
Concerning the extinction properties of normal star
forming galaxies, we are able to
interpret the observed lack of correlation between the attenuation
measured at H$\alpha$ and in the UV band as a consequence of age
selective extinction. We also find that, for these galaxies with
modest SFR, the ratio FIR/UV provides the best constraints on
the UV attenuation.
The accurate treatment of lines and continuum in dusty galaxies
also allows to deal with different SFR
estimators in a consistent way, from the UV to radio wavelengths,
and to discuss the uncertainties arising from the different physical
conditions encountered in star forming galaxies.
We provide our best estimates of SFR/luminosity calibrations, together 
with their expected range of variation. 
It results that SFR derived through H$\alpha$, even when corrected for 
extinction using the Balmer decrement, is affected by important uncertainties
due to age selective extinction. Another remarkable result is that SFR 
from UV luminosity corrected 
by means of the ratio FIR/UV has a small uncertainty. Finally, our model
provides a calibration of SFR from radio luminosity; its value 
differs from estimates from other works, but we are 
able to reproduce the observed FIR/radio ratio.
These results are relevant to estimates of the contribution of
disk galaxies to the cosmic SFR at $z \leq 1$.
 }

\maketitle

\section{Introduction}
\label{intro}

Massive stars in young stellar generations leave their signature
as UV, ionizing photons and supernova (SN) explosions. Thus,
ongoing star formation in galaxies is traced by the reprocessing
of stellar radiation (line and continuum emission from the
surrounding ionized gas, thermal emission by dust grains)
and by synchrotron emission of SN accelerated electrons. The
hydrogen recombination lines have been widely used as tracers
of the current star formation rate (SFR), because of the direct
proportionality between their intensity and the number of living
massive stars. 
In the absence of dust, nebular emission models predict an almost
constant ratio of H$\alpha$ to H$\beta$ for a large variety of 
environments (e.g. Osterbrock \cite{oste89}); so, the dust 
absorption can be estimated by comparing the observed ratio to the 
theoretical one (Balmer decrement method).

However, the discovery by IRAS of powerful and highly dust
enshrouded starbursts (Soifer et al. \cite{soif86}), namely the Luminous
and Ultra Luminous Infrared Galaxies, has underlined the necessity to 
model nebular emission by consistently taking into
account dust reprocessing. 
Moreover, an impressive wealth of information is coming from 
infrared spectral region. Both star formation and a possible
nuclear activity contribute to infrared luminosity, but with different 
emission properties (Rigopoulou et al. \cite{rigo99}).
The combination of different indicators (such as optical, near- and
mid-infrared emission lines, mid-infrared PAH bands, shape
of the continuum from the far-ultraviolet to the radio
wavelengths) should then allow to obtain a fair picture of the star
formation process and possible AGN contribution for a wide class
of active galaxies.

Two models have been recently published in which spectral
synthesis and photo-ionization codes are coupled. Charlot \&
Longhetti (\cite{char01}, henceforth CL01) combined a population 
synthesis code with the photo-ionization code CLOUDY (Ferland 
\cite{ferl01}), including an approximate prescription to 
estimate the dust absorption. According to CL01, the presence of
dust increases the uncertainty of standard SFR estimators such as
H$\alpha$ or [\ion{O}{ii}] luminosities, from a factor of a few to
several decades. Only the simultaneous consideration of
several other lines and spectral features can reduce the
uncertainty within a factor of a few. In a similar way, Moy
et al. (\cite{moyr01}, hereafter MRF01) interfaced the
evolutionary synthesis model P\'EGASE (Fioc \& Rocca-Volmerange
\cite{fioc97}) with CLOUDY. In these models the dust processes are
partially treated (extinction but not emission) using a simple
screen approximation.\footnote{Additional details on these models
are given and compared to our method in Sect. \ref{nebgalassia}.}

However, growing evidence has been collected in the recent past
showing that a physical understanding of dust effects in galaxies
requires the inclusion of different environments, arranged in a
rather complex geometry. A sophisticated treatment tends to become
more and more important as the obscuration gets higher. 

In this paper, we propose a new method to compute nebular 
emission in star forming galaxies. 
The method was implemented in the spectrophotometric synthesis model 
GRASIL\footnote{The GRASIL code and updated information on it can be
found at the URL: \\ 
{\tt http://web.pd.astro.it/granato/grasil/grasil.html}.
(Silva et al. \cite{silv98}), and then it was used to study
the properties of normal star-forming galaxies.}

The main advantage with respect to previous treatments is that
GRASIL already provides a sound treatment of all the aspects of dust
reprocessing, since it makes use of a geometry which is much closer to 
reality than
a screen between the stars and the observer. Dust modeling is
thus related to real physical parameters, describing the different
distributions of dust and stars in molecular complexes and diffuse
components, and their age dependence. 

One of its important features is that it takes into account that,
since stars are born in dense environments (the
molecular clouds) and progressively become less obscured, the
relative geometrical arrangement of dust and stars depends on the
age of the stellar generation considered. Granato et al. (\cite{gran00}) 
have shown that this age-dependent extinction can explain the
differences between the observed attenuation laws in normal and
starburst galaxies (see also Poggianti et al. \cite{pogg01}).

GRASIL has been shown to
reproduce the UV to radio continuum SEDs of galaxies, at low and
high redshift, and in different evolutionary stages (Silva et al.
\cite{silv98}, Franceschini et al. \cite{fran98}, Granato et al. 
\cite{gran00}, Rodighiero et
al. \cite{rodi00}, Granato et al. \cite{gran01}).

The paper is organized as follows. In Sect. \ref{secmodel} we
describe how we modeled stellar radiation and the computation of nebular
emission from the single \hii\ regions. Sect.
\ref{nebgalassia} explains the calculation of nebular emission
from galaxies and in the following sections 
our model is applied to the case of normal star-forming
galaxies. In particular in Sect. \ref{extinction}
the different methods to estimate the attenuation are discussed. 
In Sect. \ref{sfrestimators} different SFR
estimators, together with their uncertainties, are presented.
In Sect. \ref{secirlines} we discuss the use of IR nebular lines.
Thus, in Sect. \ref{secdiscus}
the main results of the paper are discussed, and then summarized in the 
last section. Finally, some technical details of Sect. \ref{sec_ionspe} 
and \ref{nebgalassia} are described in appendix \ref{app_an} and \ref{app_u}.


\section{Population synthesis with gas and dust}
\label{secmodel}

In this section we summarize the main features of our population
synthesis code GRASIL. More details can be found in Silva et al.
(\cite{silv98}) and Granato et al. (\cite{gran00}). GRASIL  represents 
galaxies by
means of two main components characterized by different geometries:
a spheroidal component (the bulge) and a disky component. Dust,
that may be present in both components, is divided in two phases:
i) dense molecular clouds (MCs), where star formation is active,
and ii) diffuse medium (or cirrus).

Young stars are assumed to be born into MCs, and to leave them
progressively as their age increases. As a consequence, the
fraction of light of young simple stellar populations (SSPs)
radiated inside MCs is a decreasing function of SSPs age, 
parameterized
by the ``escape time". Thus, the light of young stars will be attenuated 
by both the MCs and the cirrus, while older populations will only be
affected by dust in the cirrus component. The molecular cloud is
modeled as a thick spherical shell of dense gas (and dust) around
a central point source, representing all the stellar content of
the cloud. The time dependence of the escape fraction gives rise
to the age-selective extinction because younger stellar
generations are more attenuated than older ones.

Massive stars are supposed to ionize the surrounding medium 
and to give rise to the corresponding line and continuum 
nebular emissions. This radiation comes
generally from within MCs and it is accordingly extinguished.
However, it is worth noticing that, when the escape time is short
enough, a significant number of ionizing photons can arise from
star generations outside MCs and, consequently, we will also
consider \hii\ regions extinguished only by the cirrus
component. To compute nebular emission we have used the
photo-ionization code CLOUDY (version 94), as detailed in the
following sections.

Since our interest is focused on the star formation process and
the obscuration of star-forming regions, we do not include in our
model the emission due to an Active Galactic Nucleus (AGN). We
neglect the ionization due to UV radiation from post-AGB stars,
since in presence of even a modest star formation activity their
contribution is very low (Binette et al. \cite{bine94}). Moreover, we
neglect the contribution of shocks produced by SN explosions, that
is typically low (Kewley et al. \cite{kewl01}). Finally, we do not include
line emission from photo-dissociation regions and diffuse warm
neutral/low ionized medium, although some fine structure IR lines
are efficiently produced in these media. As a consequence their
luminosities in our model will be only lower limits; this problem
will be discussed in Sect. \ref{sec_pdr}.

The final output of our model is a complete and detailed spectrum of 
star-forming galaxies, from far-UV to the radio
wavelengths (an extension to X-ray band is in progress, Silva et al. 
in preparation), 
including stellar absorption features, nebular emission, dust and PAH 
emission.


\subsection{Stellar Radiation}
\label{secstellar}

SEDs of stellar generations (SSP, simple stellar populations)
have been computed by following the
prescriptions outlined in Bressan et al. (\cite{bres94}) and in Silva et
al. (\cite{silv98}). These SEDs cover a wide range in age and metal
content. They can be computed for an arbitrary initial mass function 
(IMF) and allow the
use of different atmosphere models, from the low-resolution (but 
covering  a wide parameter space) Kurucz-Lejeune models (Kurucz
\cite{kuru93}; Lejeune et al. \cite{leje98}), to the 
intermediate-resolution models of Pickles (\cite{pick98}) and 
Jacoby et al. (\cite{jaco84}). In the latter two
cases, which are derived from observed stars, the corresponding
fluxes have been extended into the unobserved region by means of
the Lejeune et al. (\cite{leje98}) models. The use of higher resolution
models is particularly useful when dealing with emission lines
superimposed to absorption features of the intermediate age
populations. As for the most massive stars, we have adopted the
atmospheric models by Schaerer et al. (\cite{scha96}) for mass-losing blue
supergiants, and the models by Schmutz, Leitherer \& Gruenwald
(\cite{schm92}) in the Wolf Rayet (WR) phase. It will be stressed below
that the emission from \hii\ regions has been computed in a way
that is almost independent of the detailed shape of the adopted
SEDs of stellar populations.


\subsection{\hii\ regions model}
\label{modellohii}

The emission spectrum from a single \hii\ region depends on two
main ingredients: the SED of the ionizing star cluster and the
properties of the excited gas. For a given stellar evolution
scenario, the SED is determined by the IMF and the total mass,
age and metallicity of the cluster, as well as by the adopted model
atmospheres. On the other hand, the
important properties of the excited gas are the density, the
chemical composition and the geometry. In a galaxy the situation
is more complicated and star forming regions possibly have
different ages and different metallicities. In  starbursts, star
clusters of different ages and metallicities may coexist within
the same star forming region. To compute the line emission
intensities, one should consider the spectrum of the ionizing
source, provided by the recent star formation history, and use a
photoionization code with suitable values of the gas parameters.
To reduce the computing cost, one can use a pre-built library of
line intensities, corresponding to different ages of the ionizing
SSP and different metal contents of both the SSP and the gas.
This still requires different libraries for different masses of
the ionizing SSPs, assumed IMFs and model atmospheres. It
becomes particularly time consuming in applications requiring a
large number of models.

Our approach has been to pick out the physical
parameters which actually affect the emission properties of \hii\
regions. The analysis described in the following section shows
that the emission line spectrum of an \hii\ region with fixed gas
properties (metallicity, density and geometry) is described with
reasonable precision by only three quantities: the number of
ionizing photons for \ion{H}{i}, \ion{He}{i} and \ion{O}{ii}
($Q_{\mathrm H}$, $Q_{\mhe}$, and $Q_{\mathrm O}$, defined in eq.
\ref{defqq}). This method
allows us to get rid of the particular SSP model and IMF. In
fact, different ionizing sources that provide the same values of
$Q_{\mathrm H}$, $Q_{\mhe}$, and $Q_{\mathrm O}$, will  produce the same
emission line spectra, within a reasonable accuracy.

Thus, we computed a library of photoionization models as a function
of $Q_{\mathrm H}$, $Q_{\mhe}$, and $Q_{\mathrm O}$. When estimating the
actual line emission due to a given stellar population, we compute
$Q_{\mathrm H}$, $Q_{\mhe}$, and $Q_{\mathrm O}$ from the corresponding
SED, and then interpolate the value from the above library.

The \hii\ region library can be freely retrieved through 
the GRASIL web page (see Sect. \ref{intro}).

In the following, we will describe the adopted procedure and
test its accuracy.


\subsubsection{Ionizing spectra}
\label{sec_ionspe}

In order to have photoionization models as a function of
$Q_{\mathrm H}$, $Q_{\mhe}$, and $Q_{\mathrm O}$, we approximate real
SEDs with piece-wise blackbodies, in a way such that the
space of $Q$ parameters is covered as much as possible.

The Lyman continuum spectra of typical young stellar populations
show two prominent discontinuities, at 504.1 \AA\ and 227.8 \AA ,
corresponding respectively to the ionizing energies for \ion{He}{i} and
\ion{He}{ii}. At wavelengths smaller than 227.8 \AA\
there are very few photons. 
Between the Lyman break, \ion{He}{i} break, and \ion{He}{ii}
break, spectra can be reasonably well represented by
blackbodies. Thus, when computing the libraries, we used the
following functional representation of the ionizing SEDs:
\begin{equation} 
\label{paramspe}
F_\nu= \left\{
\begin{array}{ll}
0&~~~\hbox{for } \lambda < 227.8~\mathrm{\AA}\\
A_{\mhe}B_\nu (T_{\mhe}) &~~~\hbox{for }
227.8 ~\mathrm{\AA} < \lambda < 504.1 ~\mathrm{\AA}\\
A_{\mathrm H}B_\nu (T_{\mathrm H})&~~~\hbox{for }
504.1 ~\mathrm{\AA} < \lambda < 911.76 ~\mathrm{\AA}\\
A_{\mathrm{ni}}B_\nu (T_{\mathrm{ni}})&~~~\hbox{for }
911.76 ~\mathrm{\AA} < \lambda \\
\end{array}
\right. ~,
\end{equation}
where $B_\nu (T)$ is the Planck function at temperature $T$. This
function depends on the 6 quantities $A_{\mhe}$, $T_{\mhe}$,
$A_{\mathrm H}$, $T_{\mathrm H}$, $A_{\mathrm{ni}}$ and $T_{\mathrm{ni}}$. However, we
found that the SEDs of young stellar populations can be well
approximated by writing these 6 quantities in terms of only three
parameters. These are the numbers of ionizing photons for
\ion{H}{i}, \ion{He}{i} and \ion{O}{ii} ($Q_{\mathrm H}$, $Q_{\mhe}$, 
and $Q_{\mathrm O}$):
\begin{equation}
\label{defqq} { Q_{\mathrm H}}=\int_{\nu_{\mathrm H}}^\infty \!\!
\frac{F_\nu}{h\nu} \md\nu~,~ { Q_{\mhe}}=\int_{\nu_{\mhe}}^\infty \!\!
\frac{F_\nu}{h\nu} \md\nu~,~ { Q_{\mathrm O}}=
\int_{\nu_{\mathrm O}}^\infty 
\!\!
\frac{F_\nu}{h\nu} \md\nu ~,
\end{equation}
where $\nu_{\mathrm H}$, $\nu_{\mhe}$ and $\nu_{\mathrm O}$
are the photoionization threshold frequencies for 
respectively \ion{H}{i},
\ion{He}{i}, and \ion{O}{ii} (the last one correspond to 
$\lambda_{\mathrm O}=350.7$ \AA).
The relationship between the $Q$ values and the parameters of 
analytical spectra is described in appendix \ref{app_an}.

We built a library for a grid of
values of $Q_{\mathrm H}$, $Q_{\mhe}$, and $Q_{\mathrm O}$, and for
different assumptions on the gas density, metallicity and filling
factor ($\epsilon$). When the filling factor is different from 1,
the gas in the \hii\ region is supposed to be divided in small
clumps, and $\epsilon$ is defined as the
ratio between the volume occupied by the clumps and the total
volume of the \hii\ region. The gas density refers to the hydrogen
density inside the clumps. \hii\ regions are assumed to be
spherical and ionization bounded, with a constant density along
the radius, and with a covering factor of 1. 
The abundance of elements respect to H are 
from McGaugh (\cite{mcga91}). These abundances are relative to 
gas phase, thus they already account for the fraction stored in dust 
(depletion).

The characteristics of dust inside \hii\ regions and its effects
on the emerging emission are still poorly known. 
However, it is reasonable that the main effect is that 
a fraction $f_\mathrm{d}$ (typically $\sim$ 30\%, see DeGioia-Eastwood, 
\cite{degi92}) of ionizing photons 
is absorbed by dust. To approximate the consequence, one can 
multiply the nebular emission of a dust-free model by a
factor $1-f_\mathrm{d}$.
Thus, absorption by internal dust has not been considered 
in \hii\ region models presented here. Moreover, including dust inside 
photo-ionization models would introduce new free parameters,
and the library could be coherently used only in some specific cases.

It is worth noticing that the emission properties of the ionized
gas (i.e. the ratio of different line intensities) at a fixed
density do not depend directly on the geometry of the gas, 
while it shows a dependence
on the ionization parameter $U$, that expresses the number of
ionizing photons per particle. For a thin spherical shell at
distance $R$ from a central ionizing source, the ionization
parameter is $U=Q_{\mathrm H}/(4\pi R^2 c n_{\mathrm H})$; it can 
be shown (see appendix \ref{app_u})
that the volume averaged ionization parameter $\langle U \rangle$
for a sphere, with density $n_{\mathrm H}$ and filling factor 
$\epsilon$, is
\begin{equation} 
\langle U \rangle =\frac{3\alpha_\mathrm{B}^{2/3}({\mathrm H},T)}{4c}
\left(\frac{3Q_{\mathrm H} n_{\mathrm H}\epsilon^2}{4\pi}\right)^{1/3} ~,
\label{udiq} 
\end{equation}
where $\alpha_\mathrm{B}({\mathrm H},T)$ is the recombination coefficient of
hydrogen in the case B. Thus, in our geometry $\langle U \rangle$
is proportional to $(Q_{\mathrm H} n_{\mathrm H}\epsilon^2)^{1/3}$ while
for the plane parallel geometry $U\propto Q_{\mathrm H}/n_{\mathrm H}$.

As a consequence, our results remain also valid for different
geometries and/or filling factors, provided that we renormalize
$Q_{\mathrm H}$ to maintain the  same value of $\langle U \rangle$.

The computed lines are 54 H and \ion{He}{i} recombination lines and
60 nebular line of other elements with wavelength from UV to FIR.
The complete  list of lines and the space of parameters covered by the 
library can be found in the web page.


\subsubsection{Nebular continuum emission}

The nebular continuum emission can be easily evaluated using 
the Str\"omgren theory for the case B approximation 
(see Osterbrock \cite{oste89}) and added to the
emerging spectrum.

We get
\begin{equation}
E_\nu =Q_{\mathrm H}\frac{\gamma_\nu({\mathrm H},T)}{\alpha_\mathrm{B}({\mathrm H},T)}+
Q_{\mhe}\frac{\gamma_\nu({\mhe},T)}{\alpha_\mathrm{B}({\mhe},T)}~,
\end{equation}
where $\gamma_\nu({\mhe},T)$ and $\gamma_\nu({\mathrm H},T)$ are the
total emission coefficients and $\alpha_\mathrm{B}({\mhe},T)$ and
$\alpha_\mathrm{B}({\mathrm H},T)$ are the recombination coefficients, 
respectively of helium and hydrogen. In the former coefficients we
take into account  recombination, free-free emission of \ion{H}{ii} and
\ion{He}{ii} and the Ly$\alpha$ two photons decay emission. Free-free
emission is computed as described in Bressan et al. (\cite{bres02}), 
while
other coefficients are from Burgess \& Summers (\cite{burg76}), Aller
(\cite{alle84}), Osterbrock (\cite{oste89}), Ferland (\cite{ferl80}) 
and Nussbaumer \&
Schmutz (\cite{nuss84}). The \ion{He}{iii} contribution to the nebular 
emission is very small and can be neglected.


\subsubsection{Accuracy}
\label{secacc}

Before proceeding further, we need to analyze the errors introduced
by adopting the analytic SEDs (equation \ref{paramspe}) instead of
the real SSP with the same $Q$ values.

\begin{figure}
\resizebox{\hsize}{!}{\includegraphics{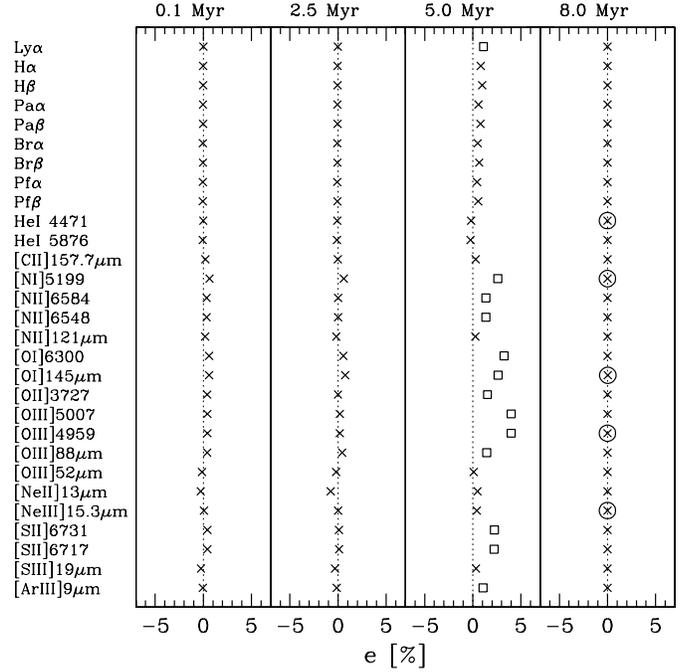}}
\caption{Differences between \hii\ region models excited
by SSP spectra with and without
high energy photons. Crosses are for difference smaller than 1\%, 
empty squares for
difference between 1\% and 5\%; circled symbols refer to very
low luminosity lines. Gas and star metallicity is 0.008; the 
stellar mass is $3\cdot 10^4$ M$_\odot$.}
\label{highen}
\end{figure}

A first problem is that our analytical spectra neglect photons
with energy higher than the \ion{He}{ii} ionization threshold
(hereafter we will refer to these photons as high energy photons).
This does not allow a realistic evaluation of the luminosity of the
\ion{He}{ii} lines. However, nebular \ion{He}{ii} lines are
superimposed to the generally more intense photospheric lines, emitted 
in the stellar atmospheres of WR stars (Conti, \cite{cont91},
Schaerer et al. \cite{scha99}). We also cannot compute line intensities 
of highly ionized elements, such as \ion{O}{iv} or \ion{Ne}{iv}.
These lines are typically either not observed, or very faint in pure
star forming galaxies, so that they are often interpreted as a signature
of AGN activity. The impact of neglecting high energy photons on
other lines is quantified by comparing the calculated emission of
\hii\ regions excited by SSP spectra with and without high energy
photons, normalized in order to have the same $Q_{\mathrm H}$. The
results (for some ages and metallicities) are shown in
Fig.~\ref{highen} for a Salpeter IMF between 0.15 and 120
M$_\odot$. The difference in the two cases is always
$<10\%$, and typically $<1\%$; the highest difference appears at
SSP ages at which the WR stars produce their maximum flux in high energy
photons. We conclude that the lack of high energy photons is not
very important, at least for the lines considered here.

\begin{figure}
\resizebox{\hsize}{!}{\includegraphics{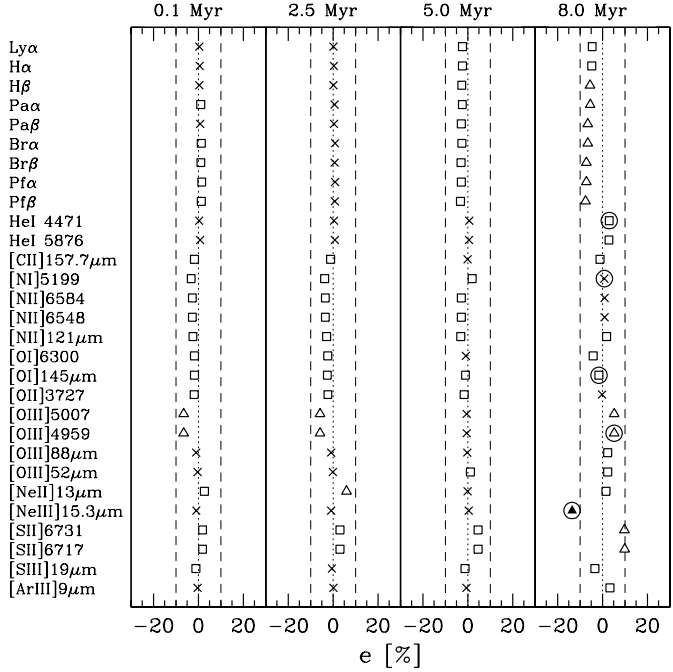}}
\caption{Differences between \hii\ region models excited by SSP 
spectra and
parametric spectra with the same $Q$s. Empty triangles represent 
differences between 5\% and 10\%,
filled triangles represent differences greater than 10\%.}
\label{cfrssp}
\end{figure}

The comparison between line intensities of \hii\ region models
excited by an SSP spectrum and those  excited by a parametric
spectrum with the same values of $Q_{\mathrm H}$, $Q_{\mhe}$, and
$Q_{\mathrm O}$ is shown in Fig.~\ref{cfrssp}. Typical differences in
the emission lines are lower than 10\%, with the larger ones
arising at lower values of $Q_{\mathrm H}$, i.e.\ older ages. Most of
them are due to discrepancies between the SSP and parametric
SEDs at wavelengths longer than the Lyman break. Indeed, photons
with wavelengths corresponding to Lyman serie lines can be
efficiently absorbed and re-emitted in optical lines. When the
ionizing flux decreases with respect to the non-ionizing one, this
effect can be important. Notice however that, when computing the
lines emitted by a combination of \hii\ regions excited by
clusters with different ages, the most important contribution
arises from the youngest populations.

As noticed in previous works (see e.g.
Stasi\'nska et al. \cite{stas01} or Rubin et al. \cite{rubi01}), the
photo-ionization models cannot be accurate when the gas
metallicity exceeds the solar value, for several reasons. As
the metallicity increases, fine structure infrared lines of metals
(that depend little on temperature) dominate the
cooling processes, but their transition probabilities have not yet been 
well determined. On the other hand, optical collisionally excited
lines, that depend strongly on the electronic temperature, become
uncertain because the temperature is regulated by infrared lines.
Furthermore, the  thermal instability of the gas affects the line
intensity when the temperature is low and the metallicity is high.
Finally, a further source of uncertainty at high metallicity  is
due to the poorly known depletion of metals into dust grains.
In order to minimize this problem, we used a set of abundances
measured by McGaugh (\cite{mcga91}) in the gas phase.

\begin{figure}
\resizebox{\hsize}{!}{\includegraphics{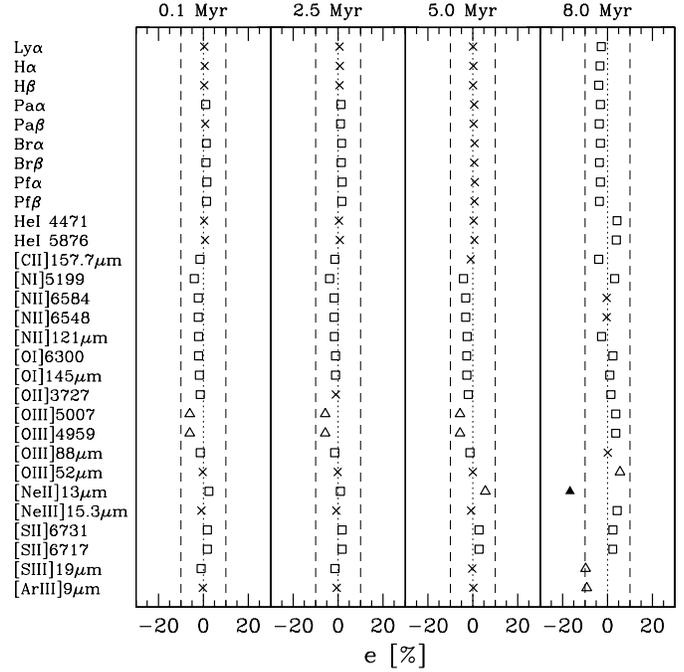}}
\caption{As Fig.~\ref{cfrssp}, but with SSPs with a top-heavy 
modified Kennicutt's IMF.}
\label{cfrimf}
\end{figure}

In order to check the flexibility of our procedure, we compare in
Fig.~\ref{cfrimf} the results obtained by repeating the above test
when adopting a very different IMF, namely a top-heavy modified
Kennicutt's IMF ($\mathrm d\log N/\mathrm d\log m=-0.4$ for 0.15~M$_\odot
<m<1$~M$_\odot$ and $\mathrm d\log N/\mathrm d\log m=-1$ for
1~M$_\odot<m<120$~M$_\odot$, Kennicutt, \cite{kenn83}). 
The differences between \hii\ region
models excited by SSP spectra and analytical ones with the same $Q$
values are not larger than in the case of the Salpeter IMF,
despite the very different slope of the IMF.

In conclusion, our analytical spectra can reproduce with a
good accuracy the nebular emission properties of \hii\
regions excited by star clusters with different IMF, age and
metallicity by only making use of the three quantities $Q_{\mathrm H}$,
$Q_{\mhe}$, and $Q_{\mathrm O}$. Obviously this accuracy could be
improved by adding some other quantity to describe the SED with more
details, but this would dramatically increase the number
of \hii\ region models used for the library.


\subsection{C and O infrared lines}
\label{sec_pdr}

\ion{C}{i}, \ion{C}{ii} and \ion{O}{i} fine structure infrared
lines are produced not only in \hii\ regions, but also in warm
neutral and ionized interstellar medium and photodissociation
regions (PDR). [\ion{O}{i}] 63.2$\mu$m and 145.5$\mu$m,
[\ion{C}{i}]369$\mu$m and 610$\mu$m infrared lines have upper
energy levels respectively of 228 K, 326 K, 62.5 K and 23.6 K 
(Kaufman
et al. \cite{kauf99}), so they can easily be produced in
neutral medium. Because carbon has a ionization potential 
(11.26 eV) which is lower
than H, the \ion{C}{ii} ion is present in PDR and in neutral
medium illuminated by far-UV stellar radiation. Indeed, the
[\ion{C}{ii}]157.7$\mu$m line is the most important coolant of
warm neutral medium. The relative contribution of
different media to these lines is still a matter of
debate (Heiles \cite{heil94}, Malhotra et al. \cite{malh01}). 
Since our model does not
include PDR or neutral gas emission, the luminosities predicted
for these lines must be taken as a lower limit.

\subsection{Comparison with observed \hii\ galaxies}
\label{diagnostici}

\begin{figure*}
\centering
\includegraphics[angle=270,width=11cm]{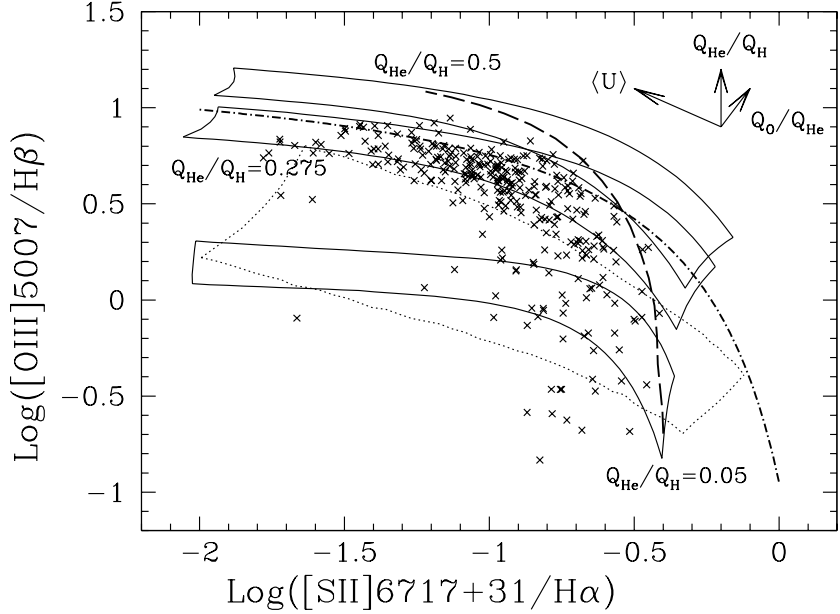}
\caption{Diagnostic diagram Log([\ion{O}{iii}]/H$\beta$) vs.
Log([\ion{S}{ii}]/H$\alpha$). Crosses are observed \hii\ galaxies
from Dessauges-Zavadsky et al. (\cite{dess00}). The three solid closed
lines show the regions occupied by \hii\  region models from our
library with three values of $Q_{\mhe}/Q_{\mathrm H}$ (0.5, 0.275
and 0.05). The arrows show shifts  at increasing  $\langle U
\rangle$, $Q_{\mhe}/Q_{\mathrm H}$, or $Q_{\mathrm O}/Q_{\mhe}$. 
Log($\langle U \rangle$) ranges from -0.867 to -3.534, and 
$Q_{\mathrm O}/Q_{\mhe}$ from 0.13 to 0.63. 
The models have a gas metallicity of
0.004, while the closed dotted line shows the position of models 
with solar metallicity and $Q_{\mhe}/Q_{\mathrm H}=0.275$. 
All the models have a
hydrogen density of 10 cm$^{-3}$. The thick long dashed line
separates the region occupied by \hii\ galaxies (on the left) from the
AGN region, as empirically defined by Veilleux \& Osterbrock
(\cite{veil87}); the thick dot-dashed line delineates the same separation
computed by Kewley et al. (\cite{kewl01}).} \label{diagra}
\end{figure*}

In Fig.~\ref{diagra} we compare our \hii\ library with observations 
of a sample of \hii\ galaxies (crosses) from Dessauges-Zavadsky et al. 
(\cite{dess00}), a revision of the Terlevich's catalogue 
(Terlevich et al. \cite{terl91}). \hii\ galaxies are typically 
defined as galaxies with spectra very similar to 
those of \hii\ regions; these objects are interpreted as 
bursting dwarf galaxies.
In the figure we show the dependence of models
on $\langle U \rangle$ (or $Q_{\mathrm H}$ via the equation
\ref{udiq}) and on the hardness of the ionizing SED, expressed by 
$Q_{\mhe}/Q_{\mathrm H}$ and $Q_{\mathrm O}/Q_{\mhe}$.

The library covers the region of the [\ion{O}{iii}]/H$\beta$ vs. 
[\ion{S}{ii}]/H$\alpha$ diagnostic diagram populated by
observed \hii\ galaxies. Some models fall in the region of the
diagnostic diagrams occupied by AGNs; these models are produced by
the hardest spectra in the library ($Q_{\mhe}/Q_{\mathrm H}\simeq
0.5$), harder than the spectra that can be built with
true SSPs.


\section{Nebular emission from a star-forming galaxy}
\label{nebgalassia}

When considering a star-forming galaxy as a whole, we need to
integrate the emission lines resulting from the effects of
different stellar populations. This cannot simply be obtained by
calculating the ionizing photon fluxes $Q_{\mathrm H}$, $Q_{\mhe} $,
and $Q_{\mathrm O}$ of the integrated spectrum and by getting the
corresponding line intensities from our interpolation tables.
Indeed, the intensity of metal lines also depends on the hardness
of the ionizing spectra and on the ionization parameter. To obtain
a more realistic description of the  nebular emission, it is then
necessary to model the formation and evolution of the population
of \hii\ regions, each one characterized by a different intensity,
hardness and ionizing parameter. This may be accomplished by
splitting the recent SF history in subsequent episodes of suitable
duration, and by computing their separate contributions to the
emission lines.

The total emission in the line $l$ ($E_l$) at the epoch of
observation $T$, may be written as
\begin{equation}
\label{emisline}
E_l=\sum_j N_{{\mathrm \hii},j}E_l^*(Q^*_{{\mathrm H},j}, 
Q^*_{{\mhe},j},Q^*_{{\mathrm O},j},Z_{\mathrm{gas}})~,
\end{equation}
where $N_{{\mathrm \hii},j}$ is the number of \hii\ regions that have
formed in the time interval $[T-t_{j+1},T-t_j]$, $Q^*_{\mathrm{H,He,O},j}$
are the corresponding ionizing photon fluxes,
$Z_{\mathrm{gas}}$ is the current metallicity of the gas and $E_l^*$ is the
emission from the single \hii\ region.

$N_{{\mathrm \hii},j}$ is obtained by assuming that each \hii\ region is
illuminated by a single cluster of total mass $M^*$,
so that
\begin{equation}
N_{{\hii},j}=\frac{\int_{t_j}^{t_{j+1}}\Psi(T-t)\md t}{M^*}~,
\label{nhii} 
\end{equation}
where $\Psi$ is the SFR.
$Q^*_{{\mathrm H},j}$ for the single \hii\ region at different 
ages can be obtained by
\begin{eqnarray}
\nonumber
Q^*_{{\mathrm H},j}&=&\int_{\nu_{\mathrm H}}^\infty 
\frac{L^*_{\nu,j}}{h\nu}\md\nu = \\
&=&
\frac{\int_{\nu_{\mathrm H}}^\infty \int_{t_j}^{t_{j+1}}\Psi(T-t)
(S_\nu(t,Z(T-t))/h\nu)\md t\md\nu}{N_{\hii,j}}
\label{qstar}
\end{eqnarray}
where we have emphasized the dependence on the metallicity of the
ionizing spectra. Similar expressions  hold for $Q^*_{{\mhe},j}$, 
and $Q^*_{{\mathrm O},j}$.

The emission properties of the $j$ population depend on
the ratios $Q^*_{\mhe,j}/Q^*_{{\mathrm H},j}$ and
$Q^*_{{\mathrm O},j}/Q^*_{\mhe,j}$, that are constrained 
by the age of the $j$ population, and on the average ionization parameter of
the \hii\ regions $\langle U \rangle^*_j$.

The correspondence between the integrated Lyman continuum of the
population $j$ (expressed by $N_{{\mathrm \hii},j}\times Q^*_{{\mathrm
H},j}$) and $\langle U \rangle^*_j$ is determined by the stellar
mass $M^*$ of ionizing clusters through equations \ref{nhii},
\ref{qstar} and
\begin{equation}
\langle U \rangle^*_j
=\frac{3\alpha_\mathrm{B}^{2/3}}{4c} \left(\frac{3Q^*_{{\mathrm H},j} 
n_{\mathrm H}\epsilon^2}{4\pi}\right)^{1/3}~.
\label{udiqstar}
\end{equation}
As shown in Fig.~\ref{diagra}, \hii\ galaxies exhibit a big
variation of the ionization parameter value, which means that the
value of stellar mass $M^*$ and/or of the filling factor are
subject to substantial variations.
Observations (e.g. Kennicutt \cite{kenn84}) suggest that $M^*$ varies
approximately from 1000 to $10^6$ M$_\odot$ and $\epsilon$
from 0.1 to 0.001.

The duration of the  subsequent episodes in which we split the SF history
should be chosen by considering the evolution of the shape of the
ionizing spectra. A lower limit is set by considering that each
\hii\ region will be illuminated by all the stars formed within
the finite formation time of a typical star cluster, that we
assume to be 1 Myr (see e.g. Fuente et al. \cite{fuen01}).

As remarked in Sect. \ref{sec_ionspe}, the presence of 
dust inside \hii\ regions can be important. Thus one can
multiply the nebular emission obtained from eq. \ref{emisline}
by a factor $1-f_\mathrm{d}$.

Finally, the emission lines are extinguished in the same 
way as the stellar populations that produced them 
(see Sect. \ref{secmodel}).

In galaxies with normal dust content, almost all the photons
in the resonant Ly$\alpha$ line
are reprocessed via two-photon decay and dust absorption and 
re-emission. In our model we assume that all Ly$\alpha$ photons decade
via two-photons emission, and then these photons are possibly absorbed by 
dust. Anyhow, their contribution to the IR luminosity amounts to 
$\leq 10\%$. Only in the case of 
very low dust content, the relative contribution of absorbed 
Ly$\alpha$ photons increases respect to the contribution from 
absorbed stellar continuum and may be significant.
A detailed treatment of the Ly$\alpha$ transfer should also include 
the effects of winds and outflows, which strongly reduce the resonant 
scattering.

We conclude this section by comparing our method with other
approaches found in literature.

Fioc \& Rocca-Volmerange (\cite{fioc97}) computed $Q_{\mathrm H}$ 
from the integrated spectrum and then analytically obtain the 
H$\beta$ line intensity. 
Other hydrogen and metals lines are then rescaled in
fixed proportions. Moy et al. (\cite{moyr01}) have improved the
previous model by coupling the integrated spectrum with CLOUDY.

A more detailed analysis of emission properties in star forming
galaxies is by CL01. These authors obtained the emission from the galaxy 
by summing the line intensities of separate SSPs of given constant 
average metallicity, weighted by the corresponding SFR.
CL01 also have a more detailed model for dust absorption
(see Charlot \& Fall \cite{char00}) that accounts for an age 
selective extinction.

Compared to MRF01, our model accounts for the coexistence
of different populations of \hii\ regions, while at variance with CL01, 
it quickly computes the line emissions for different
choices of $M^*$ and gas density. Also, the \hii\ regions library can
be used for different IMFs, and we  relax the hypothesis that the
metallicities of the excited gas and of the ionizing stellar
population are the same (but a non-negligible difference is
expected only in extreme situations). 
Finally, our library can be easily used
for other population synthesis models.


\section{Attenuation properties of normal star forming galaxies}
\label{extinction}

In star forming galaxies, the intrinsic UV luminosity is directly 
related to the star formation rate; on the other hand,
the UV flux is heavily affected by extinction. Therefore it is 
important to understand
how dust absorption affects the UV luminosity and,
as a consequence, the SFR estimates.

In this section we discuss several methods to estimate the attenuation 
in normal star forming
galaxies by considering the UV, optical emission lines and FIR 
properties.
We will show that observations require the extinction of different 
stellar populations
to vary with age, with younger populations suffering a larger extinction.

Notice that we will use the term {\it attenuation} when it refers to the 
amount of light lost at a give wavelength from a complex system 
(i.e. a galaxy) and the term {\it extinction}
for the light lost from a simple source (as a star) with dust along the 
line of sight.


\subsection{Estimation of UV attenuation}
\label{extAHAUV}

Attenuation in local star forming galaxies can be derived
in several ways.

A common method is to relate the UV attenuation to the
attenuation in H$\alpha$. Calzetti (\cite{calz97}) suggested that 
the attenuation of the stellar continuum is only a fraction
($\sim 0.44$) of the attenuation of the ionized gas. By extending this 
assumption to the UV while adopting a suitable extinction law, it is 
straightforward to obtain (at $\lambda_{\mathrm{UV}}=2000$ \AA, e.g. 
Buat et al. \cite{buat02})
\begin{equation}
A_{2000}=1.6A_{\mathrm H\alpha}~.
\label{AUVHA}
\end{equation}
The gas attenuation at H$\alpha$ is derived from the Balmer decrement:
\begin{equation}
\label{balmerdec}
A_{\mathrm H\alpha}=1.086\frac{1}{e_{\beta\alpha}-1}\ln
\left(\frac{j_{\mathrm H\beta}}{j_{\mathrm H\alpha}}\frac{L_{\mathrm 
H\alpha}}{L_{\mathrm H\beta}}\right)~,
\label{AHA}
\end{equation}
where $e_{\beta\alpha}=\tau_{\mathrm H\beta}/\tau_{\mathrm H\alpha}$ 
is derived from the adopted extinction law
and $j_{\mathrm H\beta}/j_{\mathrm H\alpha}$
is the ratio of H$\beta$ and H$\alpha$ emission coefficients
(typically assumed 2.87 at an electronic temperature of $10^4$ K, 
see Osterbrock \cite{oste89}).

An alternative method to derive the UV attenuation is to consider 
the reprocessing of star light into the infrared emission. Meurer et 
al. (\cite{meur99}) (see also Calzetti et al. \cite{calz00}) assumed 
that the energy re-emitted by dust in a galaxy is provided primarily
by the UV flux of young star populations. They related 
(eq. 10 of their paper) the ratio between 
the FIR and UV fluxes (now $\lambda=1600$ \AA), 
$F_{\mathrm{FIR}}/F_{1600}$, to the UV attenuation by:
\begin{equation}
A_{1600}=2.5\log \left( \frac{F_{\mathrm{FIR}}}{0.84F_{1600}}+1\right)~,
\label{F16}
\end{equation}
where the infrared flux $F_{\mathrm{FIR}}$ is defined as the flux in 
the [40--120] $\mu$m interval derived from the 60 and 100 $\mu$m 
IRAS bands (Helou et al. \cite{helo88}), and 
$F_{1600}=\lambda_{1600}\cdot f_{1600}$ (in W cm$^{-2}$). 
Note that this relation was derived for starburst galaxies.

In the same framework, Buat et al. (\cite{buat99}) proposed a relation 
between the UV attenuation and the
$F_{\mathrm{FIR}}/F_{\mathrm{UV}}$ ratio 
($\lambda_{\mathrm{UV}}=2000$\AA),
which is suited for normal star forming galaxies:
\begin{equation}
A_{2000}=0.466+\log\left(\frac{F_{\mathrm{FIR}}}{F_{2000}}
\right)+0.433
\left[\log\left(\frac{F_{\mathrm{FIR}}}{F_{2000}}
\right)\right]^2~.
\label{A20}
\end{equation}

Buat et al. (\cite{buat02}, hereafter B02) compared the $A_{\mathrm{UV}}$
derived from the ratio $F_{\mathrm{FIR}}/F_{\mathrm{UV}}$ (eq. \ref{A20})
with the attenuation suffered by H$\alpha$ in a sample 
(called SFG sample)
of normal star forming galaxies. The SFG sample consists of 47 spiral
and irregular galaxies in nearby clusters. They were observed 
in the UV
($\lambda_{\mathrm{UV}}=$ 2000 \AA) with the SCAP, FOCA and FAUST 
instruments (Boselli et al. \cite{bose01}), in the optical (Gavazzi et 
al. \cite{gava02}), and in the FIR by IRAS. 
The galaxies were selected to
have EW(H$\alpha$)$>$ 6 \AA\ in order to minimize the errors in 
H$\alpha$ and H$\beta$ fluxes. Galaxies with
Seyfert activity were excluded from the sample. Metallicity of the 
galaxies
in the sample ranges from $\sim Z_\odot/4$ to $\sim 2 Z_\odot$.

\begin{figure}[h]
\resizebox{\hsize}{!}{\includegraphics[angle=270]{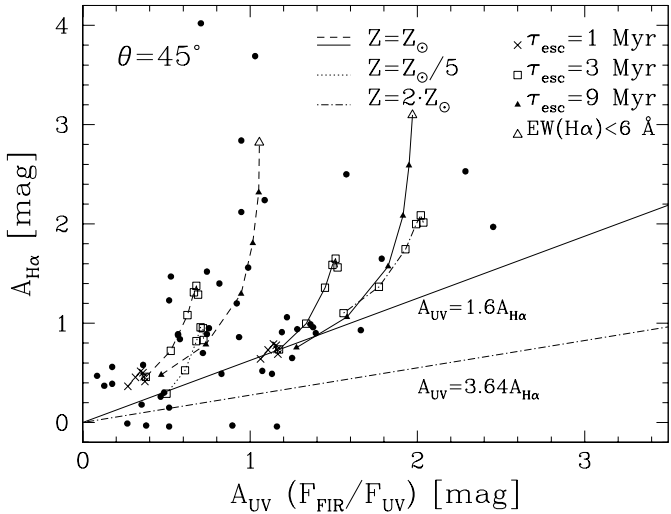}}
\caption{On the ordinate, $A_{2000}$ derived from the ratio 
$F_{\mathrm{FIR}}/F_{2000}$ (using eq. \ref{A20}); on the 
abscissa, $A_{\mathrm H\alpha}$ from Balmer decrement.
Filled circles show SFG sample data. Lines connect models 
(see Sect. \ref{sec_auv_sim}) with the 
same $M_\mathrm{G}$ and star formation history, but different 
$\tau_{\mathrm{MC}}$.
Solid lines: $M_\mathrm{G}=10^{11}$ M$_\odot$, $\nu_\mathrm{sch}=0.3$ 
Gyr$^{-1}$,
$\tau_\mathrm{inf}=$ 18 Gyr, solar metallicity; dashed lines: 
$M_\mathrm{G}=10^{10}$ 
M$_\odot$, $\nu_\mathrm{sch}=0.7$ Gyr$^{-1}$, $\tau_\mathrm{inf}=$ 
6 Gyr, solar 
metallicity. To emphasize the dependence on metallicity,
we show the first model (solid lines) for $\tau_\mathrm{esc}=3$ Myr but
$Z=0.04$ (dot-dashed line) or $Z=0.004$ (dotted line).
For simplicity only models with $\theta=45^\circ$ are shown here.}
\label{buat2}
\end{figure}

In Fig.~\ref{buat2} we report the attenuation
in H$\alpha$ versus the value of $A_{\mathrm{UV}}$ derived
from the ratio $F_{\mathrm{FIR}}/F_{\mathrm{UV}}$ for the SFG
sample, as in Fig. 2 of their paper. 
As pointed out by Buat and collaborators, the two quantities show a
lack of correlation, contrary to what is expected from eq. \ref{AUVHA}.


\subsection{$A_{\mathrm{UV}}$ in simulated normal star-forming galaxies}
\label{sec_auv_sim}

In order to give an interpretation of this observational problem,
we simulated a set of disk galaxies by exploring the space of
parameters appropriate for normal star forming galaxies. 
We computed the star formation history, gas
fraction and metal enrichment with our chemical evolution code
(Silva et al. \cite{silv98}).
In order to point out the dependence of the results on metallicity,
we considered different values  for the metallicity of stars and gas 
obtained from the chemical code.

The parameters that regulate the star formation history in our models are
the baryonic mass of the galaxy ($M_\mathrm{G}$), the gas 
infall time scale 
($\tau_\mathrm{inf}$), and the star formation efficiency $\nu_\mathrm{sch}$ of 
the assumed linear Schmidt law. The age of the galaxies has been set 
to 10 Gyr. The parameters that regulate the attenuation are the escape 
time $\tau_\mathrm{esc}$, the optical thickness of MC at 1 $\mu$m ($\tau_{\mathrm{MC}}$),
and the orientation $\theta$ of the disk galaxy with respect to the 
celestial plane  ($\theta=0^\circ$ means face-on models). 
The dust (extinction and emission) properties are similar to the 
galactic one 
(see Silva et al. \cite{silv98} for more details). The dust/gas ratio 
is assumed 
to be proportional to the metallicity.
Emission lines are computed for different gas densities $n_{\mathrm H}$, 
filling factors $\epsilon$ and stellar masses of clusters $M^*$; however
we concentrate on H recombination lines that do not depend on
$n_{\mathrm H}$, $M^*$ or $\epsilon$. 
Table \ref{parambuat} summarizes the values of the 
parameters used in our computations.

\begin{table}
\caption{Values of the relevant parameters used to model normal 
star forming galaxies. Ref. model: see Sect. \ref{sfrestimators}.}
\label{parambuat}
\centering
\begin{tabular}{lll}
\hline
\hline
Param. & All models & Ref. model\\
\hline 
$\tau_\mathrm{inf}$ & 6 -- 18 Gyr& 12 Gyr\\
$\nu_\mathrm{sch}$ & 0.3 -- 0.7 Gyr$^{-1}$ & 0.3 Gyr$^{-1}$\\
$M_\mathrm{G}$ & $10^{10}$ -- $10^{11}$ M$_\odot$& $10^{10}$  M$_\odot$\\
$Z$ & 0.004 -- 0.04 & 0.02\\
\hline
$\tau_\mathrm{esc}$ & 1 -- 9 Myr & 3 Myr\\
$\tau_{\mathrm{MC}}$ & 0.1 -- 1.25 & 0.5 \\
$\theta$ & 0$^\circ$ -- 90$^\circ$\\
$f_\mathrm{d}$ & 0.3 & 0.3\\
\hline
\end{tabular}
\end{table}

\begin{figure}
\resizebox{\hsize}{!}{\includegraphics[width=5.9cm,angle=270]{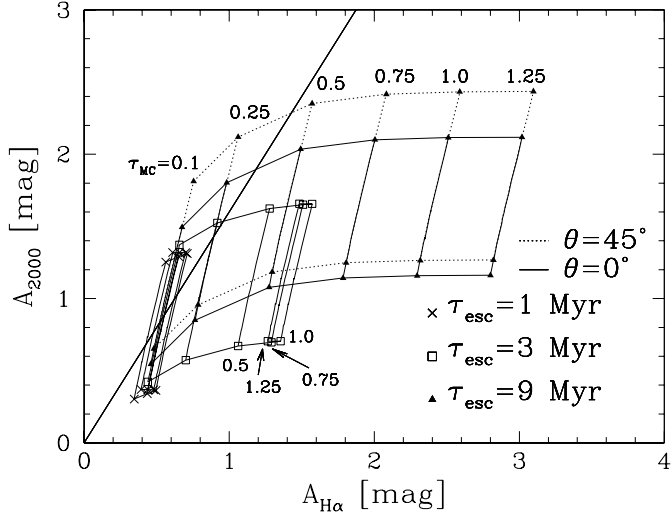}}
\caption{UV attenuation in the models versus the attenuation at H$\alpha$
derived from the Balmer decrement. Almost vertical lines connect models 
with the same optical thickness of MC ($\tau_{\mathrm{MC}}$) and escape time 
($\tau_\mathrm{esc}$), but different SF histories and dust content in cirrus. 
Only the less and most dusty 
models are associed to symbols which refer to the escape time. 
The thick solid line refers to equation \ref{AUVHA}.}
\label{uvext}
\end{figure}

The models are compared with data in Fig.~\ref{buat2};
the attenuation at H$\alpha$ is derived from the Balmer decrement 
(eq. \ref{AHA}, where we assume $e_{\beta\alpha}=1.47$), 
while the attenuation in UV has been derived by 
using eq. \ref{A20}. For simplicity, we only represent the most 
dusty (solid lines) and less dusty (dashed lines) models; all 
other cases range between the two. Models refer to a 45$^\circ$ 
inclination; face-on models show a slightly lower attenuation, 
while edge-on models have larger attenuations and scatter.

The models cover quite well the location of the
observed galaxies in this diagram and confirm a real lack
of correlation between the UV attenuation, as derived from the
$F_{\mathrm{FIR}}/F_{\mathrm{UV}}$ ratio and $A_{\mathrm H\alpha}$.

In order to clarify the origin of this scatter and to identify 
a good estimator of UV attenuation, we contrast the intrinsic UV 
attenuation -- directly extracted from our models 
($A_{\mathrm{UV}}= -2.5\log(L_{\mathrm{UV}}/L_{\mathrm{UV}0}))$ --
respectively with $A_{\mathrm H\alpha}$ from the Balmer decrement and 
$F_{\mathrm{FIR}}/F_{\mathrm{UV}}$
(see Figs.~\ref{uvext} and \ref{uvext1}).


\subsubsection{Attenuation from Balmer decrement}
\label{sec_abalmer}

The scatter in Fig.~\ref{uvext} must be entirely
ascribed to the interplay between the different stellar lifetimes
associated to the emission properties and the geometry set by the
critical escape time. In fact, H$\alpha$ is mainly produced by 
ionizing massive stars with a lifetime around 3 Myr, while UV is 
also emitted by less massive and longer-living stars. We may 
devise the following typical cases.

{\it Escape time shorter than the typical lifetime of an ionizing star}:
crosses. Independently of $\tau_{\mathrm{MC}}$, the models tend to define a 
relation  which seems however steeper than eq. \ref{AUVHA}.
This is due to the fact that both H$\alpha$ and UV emissions
are mainly produced outside MCs, so that they do not respond to 
differences in $\tau_{\mathrm{MC}}$ and the attenuation is mainly due to 
the diffuse medium.

{\it Escape time longer than the typical lifetime of an ionizing star}:
triangles. Emission lines are produced only inside MCs, so 
$A_{\mathrm H\alpha}$ essentially measures $\tau_{\mathrm{MC}}$. When 
$\tau_{\mathrm{MC}} > 0.5$ the UV flux produced inside MCs is completely 
reprocessed into the IR. Thus, $A_{\mathrm{UV}}$ saturates
while $A_{\mathrm H\alpha}$ still increases. This results in an 
almost horizontal displacement in Fig.~\ref{uvext}. 

{\it Escape time comparable to the typical lifetime of an ionizing star}:
open squares. At increasing $\tau_{\mathrm{MC}}$, the H$\alpha$ to H$\beta$ ratio
increases from the emitted value to a maximum value fixed by the H$\beta$
emitted outside the MCs and by the sum of the H$\alpha$ still coming from
within the MCs (the attenuation is lower at H$\alpha$ than at H$\beta$)
and from outside. Then the ratio decreases again to the asymptotic value
fixed by the attenuation of the diffuse gas. This causes a behavior that
is intermediate between the former two cases and, in particular, gives
rise to the turnover shown at high $\tau_{\mathrm{MC}}$. As a consequence, 
the attenuation in H$\alpha$ derived from the Balmer decrement is 
underestimated.

The variation in dust content of the galaxy and/or of the inclination, 
at constant $\tau_{\mathrm{MC}}$ and $\tau_\mathrm{esc}$, 
shifts the models in 
Fig.~\ref{uvext} along a constant direction, somewhat steeper 
than that of 
eq. \ref{AUVHA}; furthermore, a variation in metallicity corresponds 
to a variation of the amount of dust which varies the attenuation 
due to the diffuse component. These effects add further dispersion 
to the data.

Our models suggest that galaxies in Fig.~\ref{buat2} with 
$A_{\mathrm H\alpha} \geq 1.5$ mag may be characterized by escape 
times larger than the typical lifetime of the ionizing stars. Models 
with $A_{\mathrm H\alpha} \geq 2.5$ have an equivalent width 
EW(H$\alpha$) lower than 6\AA ; there are several possible explanations 
for the higher observed $A_{\mathrm H\alpha}$ with large EW(H$\alpha$),
such as a small burst (increasing the equivalent width), or an  
underestimate of H$\beta$ or [\ion{N}{ii}] lines (giving higher 
$A_{\mathrm H\alpha}$).

The results of models and observations described above point out that 
age selective extinction is present in normal star-forming galaxies 
and complicates the  picture of extinction in galaxies. The simple 
hypothesis of screen extinction does not work in this kind of galaxies; 
different ways to compute attenuation give inconsistent results 
because 
they rise from specific populations (ionizing and non-ionizing UV 
emitters) that have different lifetimes and live at different optical 
depths.


\subsubsection{Attenuation from the continuum}

A more robust estimation of the UV attenuation can be obtained 
by considering the energy reprocessed by dust in the IR.

\begin{figure}
\resizebox{\hsize}{!}{\includegraphics[angle=270]{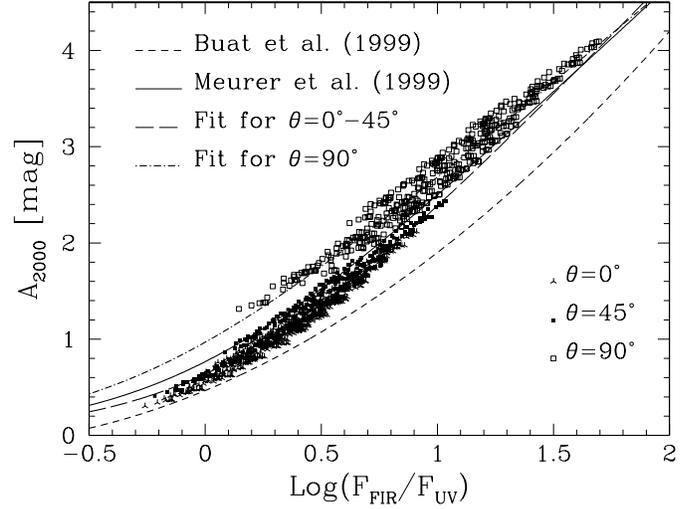}}
\caption{UV attenuation vs the ratio $F_{\mathrm{FIR}}/F_{\mathrm{UV}}$; 
lines refer to eq. \ref{F16}, eq. \ref{A20} and our fits.}
\label{uvext1}
\end{figure}

In Fig.~\ref{uvext1} we plot the intrinsic $A_{\mathrm{UV}}$ against the 
ratio $F_{\mathrm{FIR}}/F_{\mathrm{UV}}$. The solid line represents 
eq. \ref{A20} and the short dashed line refers to eq. 
\ref{F16}\footnote{To
convert attenuation at 1600 \AA\ of eq. \ref{F16} into $A_{\mathrm{UV}}$
at 2000 \AA\, we assume $F_{1600}=F_{2000}$ and 
$A_{2000}=0.9\cdot A_{1600}=0.9\cdot 2.5\log (F_{\mathrm{FIR}}/0.84
F_{2000}+1)$, as in B02.}. Models show that, contrary to the 
case of Balmer decrement, the UV attenuation correlates quite tightly 
with the $F_{\mathrm{FIR}}/F_{\mathrm{UV}}$ ratio. The origin of this 
correlation is that the FIR flux is essentially provided by the UV 
emission of young stellar populations as assumed by Meurer et al. 
(\cite{meur99}) and Buat et al. (\cite{buat99}). Thus, the 
$F_{\mathrm{FIR}}/F_{\mathrm{UV}}$ measures the ratio between the 
reprocessed energy flux and the residual energy flux in the UV. 
Outside the MCs, a contribution to the FIR comes from the optical 
emission of old stellar populations; this causes a dispersion around 
the average relation.

Similar results were obtained by Gordon et al. (\cite{gord00}); their 
models show that the 
$A_{\mathrm{UV}}$--$F_{\mathrm{FIR}}/F_{\mathrm{UV}}$
relationship is valid for different assumptions on dust properties. 
The dispersion found by Gordon et al. (\cite{gord00}) is much smaller 
than in our simulations, but in their models all the stellar populations
suffer the same extinction (i.e. no dependence on age is introduced).

Models also show that edge-on systems tend to follow a different 
relation from what is found for face-on systems; this is due 
to the greater contribution of the diffuse medium to attenuation in the 
direction of the plane of the galaxy.

Our models suggest the following relation for face-on systems:
\begin{equation}
A_{2000}=2.5\log \left(\frac{F_{\mathrm{FIR}}}{1.25F_{2000}}
+1 \right)~;
\label{A20ni}
\end{equation}
while for edge-on systems it is preferable to use
\begin{equation}
A_{2000}=2.03\log\left(\frac{F_{\mathrm{FIR}}}{0.5F_{2000}}
+1 \right)~.
\label{A20i}
\end{equation}

In Fig.~\ref{uvext1} we also compared the relations proposed by Meurer 
et al. (\cite{meur99}) (eq. \ref{F16}) and by Buat et al. (\cite{buat99})
(eq. \ref{A20}) with models. The former seems to better agree
with our model, and lies between face-on and edge-on models,
while the second tends to underestimate the UV attenuation (but only 
by 0.8 mag in the least favorable case).

Meurer et al. (\cite{meur99}) showed that the observed locus in the 
diagram
$F_{\mathrm{FIR}}/F_{\mathrm{UV}}$ vs the UV spectral index $\beta$
in starburst galaxies could be well reproduced by a screen dust model 
with increasing optical thickness. They obtained the following
relation between $A_{\mathrm{UV}}$ and $\beta$:
\begin{equation}
A_{1600}=4.43 + 1.99 \beta~.
\label{AB}
\end{equation}

Recently, Bell (\cite{bell02}) compared the spectral index $\beta$ with 
$A_{\mathrm{UV}}(\mathrm{FIR}/\mathrm{UV})$ for a sample of normal star-forming
galaxies and found that $\beta$ is  typically greater (redder)
than the expected value for starbursts.

We also found a loose $\beta-A_{\mathrm{UV}}$ correlation for disk 
galaxies. Indeed, it is worth noticing that in these galaxies the 
contribution from stars outside star forming regions is quite 
substantial and can significantly affect even the UV slope.


\section{SFR estimators}
\label{sfrestimators}

In the previous section we justified the tangled
relation between the attenuation properties of star forming galaxies
as seen at different wavelengths, by means of a model that
accounts for the complex interplay between
geometry, obscuration time and stellar lifetimes.

Using these models, we have obtained calibrations
of the SFR for a wide range of different observables
from the UV to the radio regime. Only models with 
$\tau_\mathrm{esc}\leq 9$
My are used here because the majority of
the data in Fig.~\ref{buat2} can be explained with
short escape times, $\simeq$ 3 Myr, in agreement with
other evidences coming from the analysis of the UV SEDs
and from the number counts in HR diagrams of massive stars 
(Silva et al. \cite{silv98}). In order to give a useful reference 
value of the calibration, we selected a reference model
whose  parameters are summarized in Table \ref{parambuat}.
This model is represented by a star in Fig.~\ref{buat2}. Unless 
otherwise specified, all calibrations refer to a Salpeter IMF with 
$m_\mathrm{inf}=0.15$~M$_\odot$ and $m_\mathrm{up}=120$~M$_\odot$ 
and solar metallicity.


\subsection{SFR from the continuum}

\begin{table}
\caption{Calibrations of SFR (SFR/Luminosity) from dust emission, 
in $10^{-37}$ M$_\odot$yr$^{-1}$W$^{-1}$
(or {\em a}: $10^{-33}$ M$_\odot$yr$^{-1}$W$^{-1}$\AA), and from 
UV luminosity (lower panel, in $10^{-23}$ M$_\odot$yr$^{-1}$W$^{-1}$Hz).}
\label{calibsfrfir}
\begin{center}
\begin{tabular}{lrll}
\hline
\hline
Band          & SFR/L & $\Delta$(SFR/L)\\
\hline
IR$_{8-1000}$ & 3.99  & 3.12--7.39  \\
IR            & 4.98  & 3.62--8.93  \\
FIR           & 8.82  & 6.12--16.4  \\
PAH 7.7$\mu$m & 46.7  & 31.3--81.5 {\em a} \\
ISO LW3       & 79.9  & 70.9--135. \\
MIPS 24       & 48.0  & 34.6--114. \\
MIPS 70       & 9.23  & 6.19--17.0 \\
MIPS 160      & 7.66  & 6.15--13.7 \\
\hline
UV            & 284.5  & 103.4-622.8 \\
UV$_\beta$    & 125.9  & 92.7-184.0 \\
UV$_{\mathrm{FIR}}$& 104.5  & 90.7-114.5 \\
\hline
\end{tabular}
\end{center}
\end{table}

Infrared luminosity is one of the most common SFR estimators.
The interstellar dust is able to convert in IR emission a 
substantial fraction of the UV light emitted by young stars. Thus, 
the IR luminosity is proportional to the SFR.

The relation between the SFR and the infrared luminosity is summarized in
Table \ref{calibsfrfir}. The first column indicates the observed quantity,
the second column provides the value of the calibration for the reference
model described above, and the last column provides
the range of variation of the calibration among the set of models.

The row labeled IR$_{8-1000}$ refers to the total dust emission
from 8 to 1000 $\mu$m, while IR refers to the
infrared emission estimated with the four IRAS bands (Sanders \& Mirabel 
\cite{sand96}) and FIR refers to the far-infrared emission estimated 
from the 60 and 100 $\mu$m IRAS bands (as described in Helou et al. 
\cite{helo88}).

In the following rows of Table \ref{calibsfrfir}, we show the 
calibration of the specific luminosity at the peak of the PAH emission 
feature at 7.7$\mu$m (PAH 7.7$\mu$m), the flux in the ISO band LW3, 
and the SIRTF experiment MIPS at 24$\mu$m, 70$\mu$m and 160$\mu$m.

The infrared estimators exhibit larger variations in normal 
star-forming galaxies than in starburst galaxies. Inspection of the 
above table shows that the calibrations can vary by up to 80\% around 
the reference model. In the case of normal galaxies a considerable 
part of radiation from young stars is not absorbed by MCs clouds, so that 
variations in the duration of the obscuration by MCs and 
in the optical depth of MCs produce an important change in IR emission.
Furthermore, the absorption of light from old populations by diffuse
dust may be a significant source of IR radiation, thus weakening the 
correlation with the SFR. This is quite different from more powerful 
starbursts where most of the energy produced by young stars is 
converted into IR emission, so that variations in 
absorption do not produce any important variation of the IR flux.

The ratio between SFR and UV luminosity ($\lambda_{\mathrm{UV}}$=2000 
\AA) is presented in the lower panel of Table \ref{calibsfrfir}. 
First row refers to UV uncorrected for dust extinction,
the second to the UV luminosity corrected for extinction through the 
UV slope ($A_{\mathrm{UV}}$ vs $\beta$, eq. \ref{AB} converted to 
2000 \AA ) and the last row to UV luminosity corrected for extinction 
by adopting the relation $A_{\mathrm{UV}}$ vs 
log($F_{\mathrm{FIR}}/F_{\mathrm{UV}}$) with eq. 
\ref{A20ni} and \ref{A20i}. These calibrations
refer to face-on models; calibrations from edge-on models
have a much larger range of variation, with the important exception
of the FIR-corrected estimator.

It is important to notice that the FIR-corrected UV estimator is quite 
robust; in fact, as shown in sec. \ref{sec_auv_sim}, the attenuation 
in the UV is tightly related to the ratio 
$F_{\mathrm{FIR}}/F_{\mathrm{UV}}$, for a significant variation of 
the optical depth and escape time (Fig.~\ref{uvext1}). Consequently, 
by adopting the ratio $F_{\mathrm{FIR}}/F_{\mathrm{UV}}$ to estimate 
$A_{\mathrm{UV}}$, one may get the intrinsic UV flux and
obtain a reliable SFR indicator. In summary, UV and FIR fluxes alone 
are fragile SFR indicators, while their combination is a very good 
estimator of the UV attenuation. Notice also that, because of the 
slope of the relation $A_{\mathrm{UV}}$ vs 
$F_{\mathrm{FIR}}/F_{\mathrm{UV}}$, this method is not much affected by 
uncertainties in the FIR flux due to old populations.

We also provide the dust free calibration of UV for different 
metallicities in Table \ref{calib_uvradio}.

Finally, we report in Table \ref{calib_uvradio} the new 
radio calibrations 
at 1.49 GHz and 8.44 GHz, according to Bressan et al. (\cite{bres02}), 
as a function of the stellar metallicity.
The emission at these frequencies in our model is provided by ionized gas
(thermal emission) and SN remnants (non-thermal emission).
The variation in luminosity (at constant SFR) is due to
the variation of the number of ionizing photons of SSPs with 
metallicity, of the temperature of the ionized gas for free-free 
component, and to the variation of SN rate for non-thermal contributions.
It is worth noticing that the variation of 1.49 GHz calibration with 
metallicity is $\pm$15\%.

\begin{table}
\caption{Calibrations of SFR in UV band in the dust-free case, and for radio emission 
(in $10^{-23}$ M$_\odot$yr$^{-1}$W$^{-1}$Hz) as a function of metallicity.}
\label{calib_uvradio}
\begin{center}
\begin{tabular}{cccc}
\hline
\hline
Z & UV  &  1.49 GHz &      8.44 GHz\\
\hline
0.0008      & \fan{0}78.9 &  59.5 & 166.1\\
0.004\fan{0}& \fan{0}88.7 &  65.1 & 212.6\\
0.008\fan{0}& \fan{0}94.8 &  66.6 & 230.9\\
0.015\fan{0}& \fan{0}99.7 &  67.7 & 246.2\\
0.02\fan{00}&       103.4 &  68.6 & 258.4\\
0.03\fan{00}&       107.7 &  69.8 & 269.1\\
0.04\fan{00}&       112.3 &  71.1 & 280.7\\
0.05\fan{00}&       117.4 &  72.5 & 293.5\\
\hline
\end{tabular}
\end{center}
\end{table}

Observations indicate that FIR and radio emissions are strongly 
correlated over a wide range of IR luminosities, from star-forming 
to starburst galaxies (Sanders \& Mirabel \cite{sand96}):
\begin{equation}
q=\log \frac{F_{\mathrm{FIR}}/(3.75\times 10^{12}\mbox{Hz})}{
F_{1.49\mathrm{GHz}}/(\mbox{W m}^{-2}\mbox{Hz}^{-1})}\simeq 2.35\pm 0.2~.
\label{qeq}
\end{equation}
From calibrations in Table \ref{calibsfrfir} and in Table 
\ref{calib_uvradio}, 
for the FIR and the 1.49 GHz, we obtain 
$q=2.32$ for the reference model, 
with a range of variation $q=2.05-2.48$.

Remarkably, the 0.2 dex observed scatter is fully accounted
in our models simply by the variation of the SFR-FIR  relationship 
(see also Perez-Olea \& Colina \cite{pere95}). This suggests that the 
relationship between radio emission and star formation is extremely 
tight and implies that the underlying physical mechanism
responsible for it is quite homogeneous.

Although IR and radio calibrations depend on $f_\mathrm{d}$ 
(here assumed to be 0.3), their values change by less than 5\% 
(25\% for 8.44 GHz) if $f_\mathrm{d}=0$ is assumed.


\subsection{Optical near- and mid-infrared emission lines}

H recombination lines are the only ones almost proportional
to the Lyman continuum flux and, as a consequence,
to the SFR; other emission lines also depend on other quantities so that 
their use as SFR estimators is quite dangerous.

In Table~\ref{calibsfr} we report our calibrations for some 
emission lines.
The second column shows the intrinsic calibrations (without dust);
the third column provides the calibration for the reference model 
described 
above. The forth column provides the maximum value found in our 
models (the minimum value of calibrations corresponds to the case 
without dust).

The calibrations are given assuming that the fraction of ionizing photons
absorbed by dust inside \hii\ regions ($f_\mathrm{d}$) is 0. 
Calibrations 
for different values $f_\mathrm{d}$ can be obtained by dividing the 
value reported by $1-f_\mathrm{d}$.

\begin{table}
\caption{Calibrations of SFR for emission lines 
(in $10^{-33}$M$_\odot$yr$^{-1}$W$^{-1}$, or {\it a:} in $10^{-35}$
M$_\odot$yr$^{-1}$W$^{-1}$.
No dust: the intrinsic calibrations in the case without dust.
Ref. model: calibrations for the reference model. Max value:
the maximum value found in our models. The values between brackets 
in the forth column refer to the edge-on case.}
\label{calibsfr}
\begin{center}
\begin{tabular}{llll}
\hline
\hline
Line        & No dust & Ref. model & Max value \\
\hline
H$\beta$    & 0.187 & 0.758 & 7.94  (25.8)\\
H$\alpha_0$ & 6.08  & 6.62 & 15.12 (29.3) {\it a}\\
Pa$\beta$   & 0.99 & 1.415 & 2.52  (4.37) \\
Pa$\beta_0$ & 0.99 & 0.99 & 1.06  (1.18)  \\
Pa$\alpha$  & 0.436 & 0.505 & 0.676 (0.938) \\
Br$\gamma$  & 5.69  & 6.31 & 7.86  (10.2)  \\
Br$\beta$   & 3.28  & 3.50 & 4.12  (4.96)  \\
Br$\alpha$  & 1.69  & 1.721 & 1.91  (2.10) \\
Hu$\alpha$  & 12.9 & 13.12 & 14.7 (16.2) \\
$[$\ion{O}{ii}$]$3727 & 0.127 & 0.844 & 25.2  (104.8) \\
\hline
\end{tabular}
\end{center}
\end{table}

\begin{description}

\item[H$\beta$.]
The first entry in the table is H$\beta$ uncorrected for extinction; 
this line may be the only hydrogen line detected in the spectra of 
intermediate--redshift galaxies and, in that case there is no way 
to correct it for extinction. The calibration we provide here shows 
that for a typical SFR of a few M$_\odot$/yr, the inferred
SFR may be underestimated by a factor of three.

\item[H$\alpha_0$.]
The H$\alpha_0$ row refers to the H$\alpha$ luminosity corrected
for extinction using the Balmer decrement (eq. \ref{balmerdec}). 
We notice that the decrement has
been evaluated from comparisons of the observed ratio of the
intensity of the emissions at H$\alpha$ and H$\beta$ with the expected
ratio. We did not attempt to simulate a real measure of the Balmer
decrement in a synthetic spectrum. Thus, our calibration assumes that one
is able to correct the lines for the contribution of the underlying
older populations.

It is worth noticing that, due to the age selective extinction, 
the calibration may change by a factor of two, or even more
because the extinction at H$\alpha$ is underestimated, as demonstrated 
in Sect. \ref{sec_abalmer}, when the escape time is similar to the 
lifetime of ionizing stars.

Another factor of uncertainty is provided by the orientation of the 
observed galaxy because the extinction from diffuse medium is 
characterized by a mixed geometry, which tends to produce a higher 
ratio between attenuation and reddening
than for the screen geometry.

\item[Pa$\beta$, Pa$\beta_0$.]
In Table~\ref{calibsfr} then Pa$\beta$ follows, either uncorrected 
or corrected (Pa$\beta_0$) for extinction, assuming that the intrinsic 
value of Pa$\beta$/Br$\gamma$ is 5.65 
(and $\tau_{\mathrm{Pa}\beta}/\tau_{\mathrm{Br}\gamma}=2.95$). In this 
case the extinction is lower and the SFR can be 
obtained with a large accuracy. However, it must be kept in
mind that at these wavelengths uncertainties in the photometry of
different wavebands may constitute the larger source of 
uncertainty in the
estimated extinction (eg. Calzetti et al. \cite{calz96}).

\item[Pa$\alpha$.]
The table continues with Pa$\alpha$  uncorrected for extinction. This 
line sits at the short wavelength border of the K band 
($\lambda$=18752\AA); like H$\beta$, it may
be detected by ground observations only in the spectra of intermediate 
redshift galaxies.

\item[Br$\gamma$, Br$\beta$, Br$\alpha$, Hu$\alpha$.]
The following entries in the table are Br$\gamma$, Br$\beta$, Br$\alpha$ 
and Hu$\alpha$ uncorrected for extinction. As expected, 
the extinction effects decrease when the wavelength increases,
except for Hu$\alpha$ ($\lambda=12.369\mu$m)
that falls in the absorption feature of silicates.

\item[{[\ion{O}{ii}]3727.}]
The table closes with the $[$\ion{O}{ii}$]$3727 line uncorrected for 
extinction (note we only use solar metallicity).
The importance of this line is that in high redshift galaxies
this is the only bright line that, once redshifted, remains in the 
optical domain. Recently, Jansen et al. (\cite{jans01}) investigated 
the use of this line as SFR tracer, and found that SFR estimations 
by means of $[$\ion{O}{ii}$]$3727 agree with the H$\alpha_0$
value, as long as the line intensity is corrected for extinction 
and calibrated for metallicity. However, it is not possible to perform 
these corrections without observing other lines that (if visible) are 
better estimators of SFR; therefore we conclude that
[\ion{O}{ii}]3727 is a poor indicator of SFR.

\end{description}


\section{Infrared nebular metallic lines}
\label{secirlines}

The IR spectral range contains numerous bright forbidden fine-structure 
lines that are excellent diagnostics of  gas density, hardness of the 
exciting radiation field and abundance of important elements. 


\subsection{Density and ionizing spectrum hardness diagniostics}

Different IR fine structure transitions of the same ion
have different sensitivity to collisional deexcitation.
This can be used to identify the typical electron density of the
emitting gas (Rubin et al. \cite{rubi94}).
Typical line pairs used for this purpose
are [\ion{Ar}{iii}]21.8$\mu$m/9$\mu$m, 
with a maximum sensitivity to the electron density at log($n_\mathrm{e}$) =
4.7, [\ion{Ne}{iii}]36/15.5$\mu$m (log($n_\mathrm{e}$) = 4.7),
[\ion{O}{iii}]52/88$\mu$m (log($n_\mathrm{e}$) = 2.9),
[\ion{S}{iii}]19/33$\mu$m (log($n_\mathrm{e}$) = 3.5),
and [\ion{N}{ii}]122/205$\mu$m (log($n_\mathrm{e}$) = 1.8). Though collision 
strengths for these lines may need to be reexamined (Rubin et al. 
\cite{rubi01}), it appears that they can trace the density of the 
emitting gas over a wide range, from $n_{\mathrm H}=10$ cm$^{-3}$
to $n_{\mathrm H}=10^5$cm$^{-3}$, i.e. possibly encompassing the 
typical densities of star forming regions in normal galaxies as well 
as in compact obscured starbursts (Bressan et al. \cite{bres02}).

The hardness of the ionizing spectrum and the average ionization 
parameter may be derived by analyzing ratios of line intensities 
in different ionization stages of suitable elements. Typical ratios 
are [\ion{Ne}{iii}]/[\ion{Ne}{ii}], [\ion{Ar}{iii}]/[\ion{Ar}{ii}]
and [\ion{S}{iv}]/[\ion{S}{iii}] (e.g. Giveon et al. \cite{give02}). 
Our model can be used to obtain information on the age of the 
ionizing stellar population and on the ionization parameter in 
single \hii\ regions or in starbursts of short duration.
In other conditions, this method will only provide
the average values for these quantities. We will explore these topics
in a forthcoming paper.


\subsection{Abundance determinations}

It is straightforward to extend 
to the infrared the method used for abundance determinations
for optical lines. This method is 
based on the ratio between the sum of the intensities
of the most abundant ions of an element, and that of a hydrogen 
recombination  line (typically the Br$\alpha$ line). It works
as long as the current density is lower than the critical
density for collisional deexcitation.

In the near- and mid-infrared the elements to which we can apply the 
above method are neon ([\ion{Ne}{iii}]15.5$\mu$m and 
[\ion{Ne}{ii}]12.8$\mu$m), argon ([\ion{Ar}{iii}]9$\mu$m and 
[\ion{Ar}{ii}]7$\mu$m) and sulphur ([\ion{S}{iv}]10.5$\mu$m, 
[\ion{S}{iii}]18.7$\mu$m and [\ion{S}{ii}]1.03$\mu$m). Suitable
calibrating relationships can be found in Giveon et al. (\cite{give02}) 
and Verma et al. (\cite{verm03}).

In the far-infrared the method is applicable only to nitrogen 
([\ion{N}{ii}]122$\mu$m and [\ion{N}{iii}]57$\mu$m). For these lines 
we have derived the following calibrations with respect to the 
Br$\alpha$ line (for different assumptions on the electronic 
temperature $T_\mathrm{e}$):
\begin{equation}
\frac{\mathrm N}{\mathrm H}=\left\{
\begin{array}{r}
1.907\cdot 10^{-5} \left(L_{[\ion{N}{ii}]122}+0.247 
L_{[\ion{N}{iii}]57}\right)/
L_{\mathrm{Br}\alpha} \\ \hbox{for }T_\mathrm{e}=10000\hbox{ K}\\
\\
3.546\cdot 10^{-5} \left(L_{[\ion{N}{ii}]122}+0.231 
L_{[\ion{N}{iii}]57}\right)/
L_{\mathrm{Br}\alpha} \\ \hbox{for }T_\mathrm{e}=5000\hbox{ K}\\
\end{array}
\right. 
\label{Ncalib}
\end{equation}
where N/H is the nitrogen abundance respect to hydrogen. The relevant
transition data have been taken 
from Lennon \& Burke (\cite{lenn94}), Peng \& Pradhan (\cite{peng95}) 
and Storey \& Hummer  (\cite{stor95}).

\begin{figure}
\resizebox{\hsize}{!}{\includegraphics[angle=270]{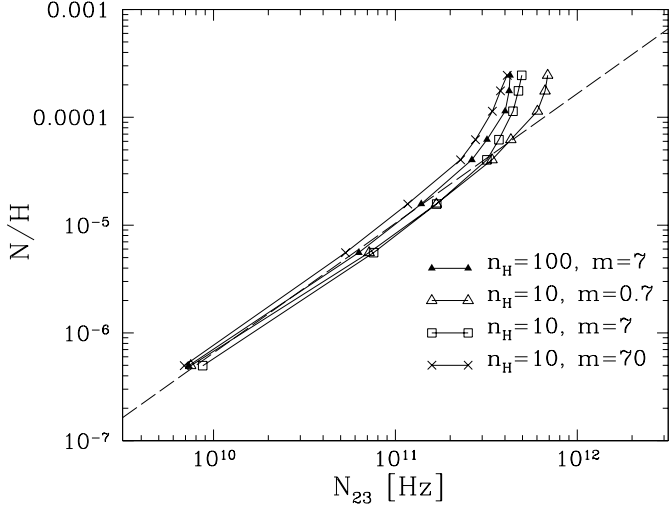}}
\caption{The relationship between nitrogen abundance (relative to 
hydrogen) and $N_{23}$ (defined by eq. \ref{equn23}) for models with 
different densities and ionization parameters. The dashed line shows a 
linear fit to models with $Z < 0.015$.}
\label{znitro}
\end{figure}

It is worth noticing that this method requires the measurement of 
the intrinsic intensity of a hydrogen recombination line. 
Thus, it is not 
applicable when these lines are not accessible, either because of 
instrumental limitation or for very high
extinction. In this case,  we suggest to adopt another indicator 
tightly correlated with the intrinsic intensity of hydrogen
recombination lines or, equivalently, with the SFR. The best SFR 
indicator we have found is the radio luminosity (Table \ref{calib_uvradio}). 
To this purpose, we have run a set of models varying the metallicity 
of the gas (from Z=0.0008 to Z=0.05), the hydrogen density, from 10 
to 100 cm$^{-3}$, and the ionization parameter. The latter is a 
function of the mass of ionizing clusters $M^*$, and of the filling 
factor $\epsilon$ through eq. \ref{udiqstar}. We stress that the 
emission lines resulting from quiescent star forming galaxies are the 
sum of the emission of \hii\ regions of different ages,
so that there is not a single ionization parameter $U$ at work. However, 
since all the ionization parameters scale linearly with the quantity
$m\equiv (M^*\epsilon^2)^{1/3}$, we have changed $m$ from 0.7 to 70
$M_\odot^{1/3}$ to simulate a similar range of variation of $U$.
The fraction of ionizing photon absorbed by dust was assumed to be 
$f_\mathrm{d}=0.3$. 

Fig.~\ref{znitro} shows the relationship in our models between the 
nitrogen abundance and the ratio $N_{23}$ defined as 
\begin{equation}
N_{23}=\left( L_{[\ion{N}{ii}]122} +0.247 L_{[\ion{N}{iii}]57}\right)/
L_{1.49\mathrm{GHz}}~.
\label{equn23}
\end{equation}
We notice that the metallicity 
is well determined for values below solar. For higher values
the relation steepens considerably making the determination
of the metal content quite uncertain. 

A linear fit to models in which the metallicity is lower than 
0.015 is plotted in Fig.~\ref{znitro}; its analytical expression is
\begin{equation}
\log \left(\frac{\mathrm N}{\mathrm H}\right)= -18.21 + 1.203 \log N_{23}~,
\label{Nfit}
\end{equation}
where $N_{23}$ is given in hertz.

We stress that
this relation holds as long as the density is below the critical density
for collisional de-excitation of the [\ion{N}{ii}]122$\mu$m transition
($n_\mathrm{e}\sim 300$ cm$^{-3}$).

A similar calibration could be obtained by adopting the FIR luminosity
instead of the radio luminosity. However, in the case of
disk galaxies the relation between the FIR luminosity and the SFR is not
as tight as that between the radio luminosity and the SFR (see
Table \ref{calibsfrfir}). This would introduce a significant scatter
in the calibration. 


\section{Discussion}
\label{secdiscus}

The first application of the model presented in this work is the analysis 
of attenuation in quiescent star-forming galaxies. 

These galaxies exhibit a poor correlation between the attenuation in 
the UV band and H$\alpha$ (B02, see also Fig.~\ref{buat2}), 
with the former generally lower than what expected from the latter
on the basis of the simple law $A_{\mathrm{UV}}=1.6A_{\mathrm H\alpha}$ 
(eq.~\ref{AUVHA}). 
Our model explains this poor correlation in the context of age selective 
extinction. As the extinction in the UV is very high, the contribution 
to the observed UV flux from stars outside the molecular clouds is 
important even for relatively small optical thicknesses ($\tau_{\mathrm{MC}}$).
It is worth noticing that the fraction of young UV emitting stars 
outside MCs increases for decreasing escape time, while very young stars 
(which ionize the gas) spend most of their life inside MCs.
Therefore, above a threshold value for $\tau_{\mathrm{MC}}$ which depends on the 
escape time, an increase of $\tau_{\mathrm{MC}}$ produces an attenuation in 
H$\alpha$ larger than that in UV.

Thus, neither observations nor modeling support the assumption of a 
constant relationship between the attenuation suffered by the continuum 
and the attenuation for the gas, at least for a current SFR 
smaller than 10 M$_\odot$/yr. As a consequence, the results of Calzetti 
(\cite{calz97}) ($A_{6563}^\mathrm{stars}/A_{\mathrm H\alpha}^{\mathrm{gas}} 
\sim 0.44$) cannot be extrapolated from starbursts to
disk star-forming galaxies.

Moreover, the extrapolation of the H$\alpha$ attenuation to the UV 
through the previous reported simple law yields UV corrected fluxes 
larger than those expected from H$\alpha$ corrected fluxes, 
as found by Sullivan et al. (\cite{sull00}, \cite{sull01}).
These authors explain the result by introducing star formation histories
that change rapidly with time
and/or a more complex model of extinction. 
On the other hand, our model explains this result as a natural 
consequence of the age selective extinction.

An additional interesting issue is that the attenuation in 
H$\alpha$ derived from the Balmer decrement can be underestimated, 
since in disk galaxies the escape time and the lifetime of very 
massive stars is often similar.

The comprehensive treatment of nebular and continuum emission allows 
to conclude that the assumption of a constant extinction for all 
stellar populations is not satisfactory.
However, the detailed models presented in this work indicate that 
$A_{\mathrm{UV}}$ can be accurately estimated
by using the ratio $F_{\mathrm{FIR}}/F_{\mathrm{UV}}$ (cfr. 
eqs.~\ref{A20ni} and \ref{A20i}). This conclusion holds 
provided that the main contribution to FIR comes from the absorbed UV 
radiation of young stars, but it does not depend on the escape times 
and on the extinction properties of dust.
The relationships we obtain are intermediate between those 
by Buat et al. (\cite{buat99}) and by Meurer et al. (\cite{meur99}), 
the latter derived for starburst galaxies.

By using our set of simulated disk galaxies, we calibrated different SFR estimators,
collected in Tables \ref{calibsfrfir}, \ref{calib_uvradio} and 
\ref{calibsfr}. Some calibrations may be compared with 
values obtained in the literature. In particular we compare the results 
with Kennicutt (\cite{kenn98}) (UV, IR and H$\alpha$) and Haarsma et al. 
(\cite{haar00}) (radio); 
these comparisons are summarized in Table \ref{conf_calib}, after a small 
correction factor (1.16),
which accounts for the slightly different IMF adopted, is applied to our estimates. 

\begin{table}
\caption{Comparison of some calibrations of SFR with results from previous
works. Calibrations are referred to a Salpeter IMF between 0.1 and 
100 M$_\odot$. Reference: {\em a:} Kennicutt (\cite{kenn98}), 
{\em b:} Haarsma et al. (\cite{haar00}).}
\label{conf_calib}
\begin{center}
\begin{tabular}{llll}
\hline
\hline
              & This work & others             & \\
UV       &1.20&$1.4\cdot 10^{-21}$ 
M$_\odot$yr$^{-1}$W$^{-1}$Hz & {\em a} \\
IR$_{8-1000}$&4.63&$4.5\cdot 10^{-37}$ 
M$_\odot$yr$^{-1}$W$^{-1}$  & {\em a} \\
H$\alpha$ &7.05&$7.9\cdot 10^{-35}$ 
M$_\odot$yr$^{-1}$W$^{-1}$  & {\em a} \\
1.49 GHz &79.6&$123.3\cdot 10^{-23}$ 
M$_\odot$yr$^{-1}$W$^{-1}$Hz & {\em b} \\
\hline
\end{tabular}
\end{center}
\end{table}

There is a good agreement for H$\alpha$, UV and IR calibrations 
(note that Kennicutt \cite{kenn98} uses the notation FIR 
for the total dust 
emission between 8 and 1000 $\mu$m), while our calibration for the radio 
emission is a factor 1.5 smaller than that proposed by Condon 
(\cite{cond92}) and Haarsma et al. (\cite{haar00}),
and a factor 1.3 larger than the one quoted by Carilli (\cite{cari02}).

It is worth noticing that our calibrations give a ratio (eq.~\ref{qeq}) 
between the radio and FIR emissions, $q=2.32$, very close to 
the observed 
value, and that we can explain in a natural way the scatter 
around the observed relation. 
In disk galaxies, this scatter is due to the variation
of the fraction of stellar emission absorbed (and re-radiated) by dust.

In the case of normal star forming galaxies, the model evidences that 
H$\alpha$, UV and even IR estimators of SFR are affected by important 
scatters. For what concerns H$\alpha$, this is due to the age selective 
extinction, that can produce an underestimate of the attenuation when 
it is derived by the Balmer decrement. 
Also IR emission by itself does not provide an accurate estimate of the SFR
in disk galaxies, because the fraction of UV radiation not absorbed by the 
dust can be high. The combination of UV and FIR luminosities
provides a very good SFR estimator essentially because, even for a modest
SFR, one recovers in the FIR what is lost in the UV.

Finally we have also discussed the utility of IR nebular lines as 
diagnostic tools for deriving the average properties of the environment 
such as number density, hardness of ionizing spectrum and gas 
metallicity. In particular we have provided a new calibration for the 
nitrogen abundance (N/H) as a function of the intensities of the 
[\ion{N}{ii}]122$\mu$m and [\ion{N}{iii}]57$\mu$m, and Br$\alpha$
lines. When the latter line is missing  or useless (e.g. when affected by
strong extinction), we suggest the possible use of the radio luminosity as
indicator of the ionizing flux $Q_{\mathrm H}$, that enters implicitely 
the above calibration. We provide a new calibration of the metallicity with
the ratio $N_{23}$ defined by equation \ref{equn23}. This new calibration 
will turn out to be particularly useful for the Herschel experiment.


\section{Summary}

Here we summarize the main points of this paper.

\begin{enumerate}
\item
We have introduced the nebular emission calculations from CLOUDY into our
spectrophotometric code GRASIL. The method which interfaces
nebular emission computations with population synthesis is based on 
pre-computed libraries of \hii\ region models. These libraries can be 
retrieved from the GRASIL web 
site\footnote{\tt http://web.pd.astro.it/granato/grasil/grasil.html}
and easily used for other population synthesis codes.

\item 
As a result, we can model the spectra of star-forming galaxies from 
far-UV to radio wavelengths, including stellar absorption features, 
nebular emission, dust and PAH emission.

\item
We applied the model to study the attenuation in normal star forming 
galaxies. The poor correlation between the extinction at UV and
H$\alpha$  found by B02 is interpreted in our model as a natural 
consequence of age selective extinction.

\item
Conversely, we found that $A_{\mathrm{UV}}$ correlates quite tightly 
with the ratio $F_{\mathrm{FIR}}/F_{\mathrm{UV}}$. 

\item
We present new calibrations for the SFR using lines and continuum from UV to 
radio wavelengths, 
and study the reliability of each estimator.
In particular, NIR H recombination lines and 
radio luminosities are very accurate SFR estimators.

\item
Although UV--, H$\alpha$-- (also corrected for extinction) and IR--derived SFR 
are plagued by significant uncertainties, the UV luminosity corrected by using the 
ratio $F_{\mathrm{FIR}}/F_{\mathrm{UV}}$  is a robust SFR estimator.

\item
We reproduce the observed value of the $q$ ratio between radio and FIR emission. 
Its scatter
is ascribed to the variation of extinction between different objects.

\item
Finally, we discuss several methods and provide new calibrations
for the determination of metal abundances by means of infrared emission
lines.
\end{enumerate}


\begin{acknowledgements}
We thank A. Boselli, V. Buat and M. Magliocchetti for enlightening discussions, and the 
referee (M. Fioc) for the very accurate scrutiny of the paper and useful suggestions.
A. B. acknowledges warm hospitality by INAOE (MEX).
We are also grateful to G. J. Ferland for the public access to CLOUDY.
This research was partially supported by the Italian Ministry for University and
Research (MIUR) and ASI.
\end{acknowledgements}


\appendix
\section{Analytical spectra}
\label{app_an}

Here we present the relationships used to describe the parameters of 
analytical spectra introduced in eq. \ref{paramspe} as functions of 
$Q_{\mathrm H}$, $Q_{\mhe}$, and $Q_{\mathrm O}$.

$T_{\mhe}$ is implicitly given by:
\begin{equation}
\frac{\int_{\nu_{\mathrm O}}^\infty F_\nu \md\nu/h\nu}{\int_{\nu_{\mhe}}^\infty 
F_\nu \md\nu/h\nu}
=\frac{\int_{\nu_{\mathrm O}}^{\nu_{\mhe^+}} B_\nu(T_{\mhe})\md\nu/h\nu}{
\int_{\nu_{\mhe}}^{\nu_{\mhe^+}} B_\nu(T_{\mhe})\md\nu/h\nu}
=\frac{Q_{\mathrm O}}{Q_{\mhe}}~,
\end{equation}
where $\nu_{\mhe^+}$ is the photoionization threshold frequency for 
\ion{He}{ii}. Once $T_{\mhe}$ is known, $A_{\mhe}$ is obtained from
\begin{equation}
A_{\mhe}\int_{\nu_{\mhe}}^{\nu_{\mhe^+}}\frac{B_\nu(T_{\mhe})}{h\nu}\md\nu=
Q_{\mhe}~.
\end{equation}

\begin{figure}
\resizebox{\hsize}{!}{\includegraphics[angle=270]{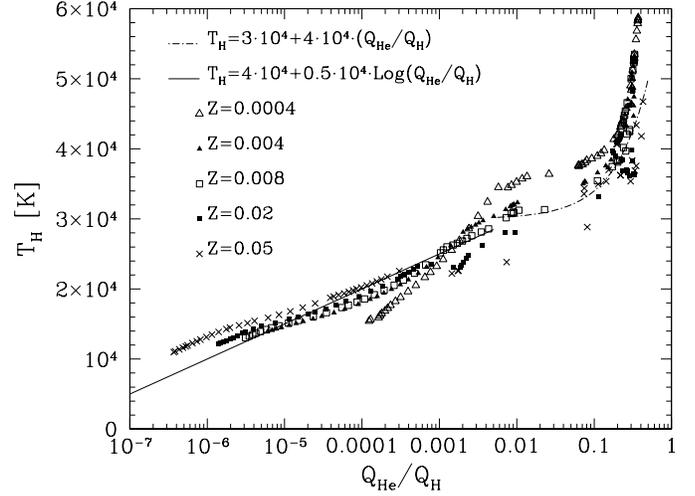}}
\caption{Temperature of ionizing continuum $T_{\mathrm H}$ as a function of
$Q_{\mhe}/Q_{\mathrm H}$ for SSPs of different age and metallicity 
(points). The lines refer to the analytical fit. (eq. \ref{eqth}).}
\label{tempqhe}
\end{figure}

Furthermore, to compute $T_{\mathrm H}$, we notice that there exists 
(for a wide range 
of age and metallicities) a correlation between $T_{\mathrm H}$ 
and $Q_{\mhe}/Q_{\mathrm H}$ of the
SSP spectra, (Fig.~\ref{tempqhe}):
\begin{equation}
\label{eqth}
T_{\mathrm H}=
\left\{
\begin{array}{ll}
3\cdot10^4+4\cdot 10^4 \frac{Q_{\mhe}}{Q_{\mathrm H}} &
\hbox{for }\frac{Q_{\mhe}}{Q_{\mathrm H}}>0.005\\
4\cdot10^4+0.5\cdot 10^4\cdot \log \left(\frac{Q_{\mhe}}{Q_{\mathrm H}}
\right)&
\hbox{for }\frac{Q_{\mhe}}{Q_{\mathrm H}}<0.005\\
\end{array}
\right.~.
\end{equation}

We use this equation to derive $T_{\mathrm H}$ for given values of 
$Q_{\mhe}/Q_{\mathrm H}$. Then $A_{\mathrm H}$ follows from
\begin{equation}
A_{\mathrm H}\int_{\nu_{\mathrm H}}^{\nu_{\mhe}}
\frac{B_\nu(T_{\mathrm H})}{h\nu}\md\nu=Q_{\mathrm H}-Q_{\mhe}~.
\end{equation}

As for the values of $A_{\mathrm{ni}}$ and of the temperature $T_{\mathrm{ni}}$, 
we notice that the region of the spectrum with $\lambda > 912$ \AA\ 
becomes important only for
relatively low ionizing fluxes. We thus simply obtained their values
after minimization of the differences between the lines
computed with full SSPs spectra and analytical spectra:
\begin{equation} 
T_{\mathrm{ni}}= \left\{
\begin{array}{ll}
{\mathrm{Max}}(T_{\mathrm H},40000~\hbox{K})& ~~ \hbox{for } 
Q_{\mhe}/Q_{\mathrm H} > 0.005\\
30000~\hbox{K} & ~~ \hbox{for }Q_{\mhe}/Q_{\mathrm H} < 0.005\\
\end{array}
\right.~,
\end{equation}

\begin{equation}
A_{\mathrm{ni}}=
\left\{
\begin{array}{ll}
2.5 A_{\mathrm H}&~~~~ \hbox{for } Q_{\mhe}/Q_{\mathrm H} > 0.005\\
3.5 A_{\mathrm H}&~~~~ \hbox{for }Q_{\mhe}/Q_{\mathrm H} < 0.005\\
\end{array}
\right.~.
\end{equation}

\begin{figure}
\resizebox{\hsize}{!}{\includegraphics{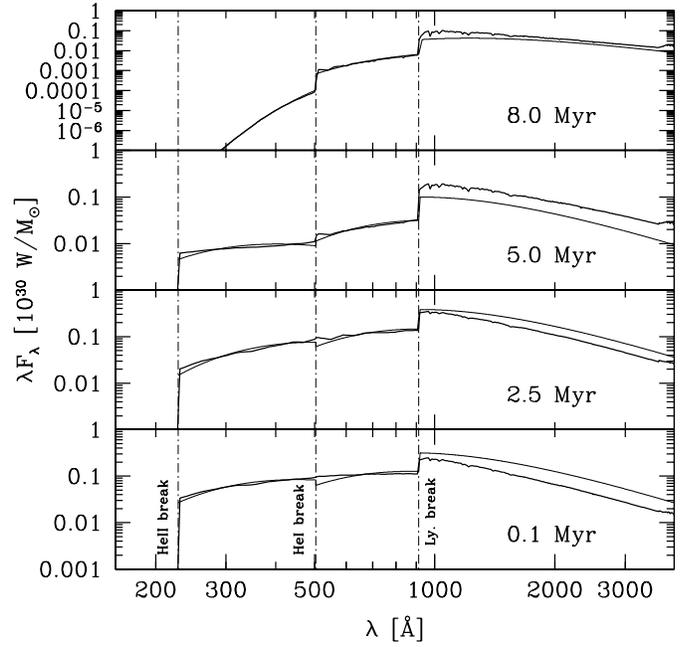}}
\caption{UV and ionizing spectra for SSPs (Z=0.008) of different ages
(thick lines) compared with the corresponding analytical spectra 
(thin lines) that have the same $Q$ values than the SSP spectra.}
\label{sed}
\end{figure}

In Fig.~\ref{sed} we compare some SSP spectra with the corresponding 
analytical spectra. These comparisons suggest that analytical spectra
can represent the Lyman continuum with reasonable precision.
However, we stress that our goal is to get line emissions computed
with analytical spectra which are as similar as possible to the ones 
computed with SSP. 

\section{Average ionization parameter for a sphere}
\label{app_u}

The volume averaged ionization parameter for a shere is 
defined as:
\begin{equation}
\label{u_aver}
\langle U \rangle =\int_0^{R_\mathrm{S}}\frac{U(r)4\pi r^2 \md r}{\frac{4\pi}{3}R^3_\mathrm{S}}~,
\end{equation}
where $R_\mathrm{S}$ is the Str\"omgren radius and $U(r)$ is the ionization parameter
of the gas at the distance $r$ from the source. 
$U(r)$ is defined by:
\begin{equation}
U(r)=\frac{Q(r)}{4\pi r^2 n_\mathrm{H}c}~,
\end{equation}
where $Q(r)$ is the number of ionizing photons per unit time that
cross a spherical surphace at distance $r$.
The variation of $Q(r)$ with the radius is given by:
\begin{equation}
\frac{\md Q}{\md r}=-4\pi r^2 n^2_\mathrm{H} \alpha_\mathrm{B}(\mathrm{H})\epsilon~.
\end{equation}
Integrating this equation we get:
\begin{equation}
\label{q_di_r}
Q(r)=Q_\mathrm{H}-\frac{4\pi}{3}r^3 n^2_\mathrm{H} \alpha_\mathrm{B}(\mathrm{H})\epsilon~.
\end{equation}
The Str\"omgren radius is defined as:
\begin{equation}
\label{stom_rad}
R_\mathrm{S}=\left(\frac{3Q_\mathrm{H}}{4\pi n^2_\mathrm{H} 
\alpha_\mathrm{B}(\mathrm{H})\epsilon}\right)^{\frac{1}{3}}~.
\end{equation}

By substituting eqs. \ref{q_di_r} and \ref{stom_rad} in eq. 
\ref{u_aver}, we get:
\begin{eqnarray}
\nonumber
\langle U \rangle &=&
\int_0^{R_\mathrm{S}} \left( \frac{Q_\mathrm{H}}{4\pi r^2 n_\mathrm{H} c}- 
\frac{r n_\mathrm{H} \alpha_\mathrm{B}(\mathrm{H})\epsilon}{3c} \right)
\frac{3r^2\md r}{R^3_\mathrm{S}}=\\
\nonumber
&=&\frac{3Q_\mathrm{H}}{4\pi  n_\mathrm{H} c}\left(\frac{4\pi n^2_\mathrm{H} 
\alpha_\mathrm{B}(\mathrm{H})\epsilon}{3Q_\mathrm{H}}\right)^{\frac{2}{3}}-\\
\nonumber
&-&\frac{n_\mathrm{H} \alpha_\mathrm{B}(\mathrm{H})\epsilon}{4c}
\left(\frac{3Q_\mathrm{H}}{4\pi n^2_\mathrm{H} 
\alpha_\mathrm{B}(\mathrm{H})\epsilon}\right)^{\frac{1}{3}}=\\
&=&\frac{3}{4c}\left(\frac{3Q_\mathrm{H}
\alpha_\mathrm{B}^2(\mathrm{H})n_\mathrm{H}\epsilon^2}{4\pi}\right)^{\frac{1}{3}}~.
\end{eqnarray}


\end{document}